\documentclass[3p,times,twocolumn]{elsarticle}

\usepackage{ecrc}

\volume{00}

\firstpage{1}

\journalname{Nuclear Physics B Proceedings Supplement}

\runauth{Graeme Watt}

\jid{nuphbp}

\jnltitlelogo{Nuclear Physics B Proceedings Supplement}

\usepackage{amssymb}
\usepackage[figuresright]{rotating}

\begin{document}

\begin{frontmatter}



\dochead{}

\title{MSTW PDFs and impact of PDFs on cross sections at Tevatron and LHC\tnoteref{label1}}
\tnotetext[label1]{To appear in the proceedings of the Ringberg Workshop on ``New Trends in HERA Physics 2011'', Ringberg Castle, Tegernsee, Germany, 25--28 September 2011.  Report no.:  CERN-PH-TH/2011-315.}


\author{Graeme Watt}

\address{Theory Group, Physics Department, CERN, 1211 Geneva 23, Switzerland}

\begin{abstract}
  We briefly summarise the ``MSTW 2008'' determination of parton distribution functions (PDFs), and subsequent follow-up studies, before reviewing some topical issues concerning the PDF dependence of cross sections at the Tevatron and LHC.  We update a recently published study of benchmark Standard Model total cross sections ($W$, $Z$, $gg\to H$ and $t\bar{t}$ production) at the 7 TeV LHC, where we account for all publicly available PDF sets and we compare to LHC data for $W$, $Z$, and $t\bar{t}$ production.  We show the sensitivity of the Higgs cross sections to the gluon distribution, then we demonstrate the ability of the Tevatron jet data, and also the LHC $t\bar{t}$ data, to discriminate between PDF sets with different high-$x$ gluon distributions.  We discuss the related problem of attempts to extract the strong coupling $\alpha_S$ from only deep-inelastic scattering data, and we conclude that a direct data constraint on the high-$x$ gluon distribution is required to obtain a meaningful result.  We therefore discourage the use of PDF sets obtained from ``non-global'' fits where the high-$x$ gluon distribution is not directly constrained by data.
\end{abstract}


\end{frontmatter}

% \linenumbers


\section{Introduction}
\label{sec:introduction}

The parton distribution functions (PDFs) of the proton are a non-negotiable input to almost all theory predictions at hadron colliders.  The proton PDFs are determined by several groups from (global) analysis of a wide range of deep-inelastic scattering (DIS) and related hard-scattering data.  The DIS data from HERA are perhaps the single most important input to global PDF fits.  It is even possible to extract PDFs based \emph{only} on HERA DIS data, albeit at the expense of leaving some kinematic regions and flavour combinations unconstrained, or alternatively by imposing a severe parameterisation constraint.  The HERA data are therefore generally supplemented with DIS and Drell--Yan data from fixed-target experiments.

Of course, we need to know PDFs to predict Tevatron and LHC cross sections, but this argument can also work the other way around.  If a cross section at a hadron collider is predicted with relatively small theoretical uncertainty, and it is sensitive to PDFs in a kinematic region poorly constrained by HERA and fixed-target data, then precise measurements of hadron collider cross sections can give important information on PDFs.  To give an example, inclusive jet production at the Tevatron is currently essential to directly constrain the high-$x$ gluon.

The contents of this contribution to the proceedings are as follows.  First in Sec.~\ref{sec:mstw} we briefly review the status of the ``MSTW'' determination of PDFs.  However, this write-up will mostly be based on two recent papers~\cite{Watt:2011kp,Thorne:2011kq}, with some updates to account for new PDF sets and LHC data released subsequent to their publication.  To avoid a deluge of plots we will only present a limited selection in this write-up, and a more extensive collection can be found at a public webpage~\cite{mstwpdf}.  In Sec.~\ref{sec:bench} we describe a recent benchmark exercise and give the status of the most recent PDF sets from all fitting groups (as of December 2011).  In Sec.~\ref{sec:WandZ} we discuss $W$ and $Z$ production at the LHC, and the sensitivity to the quark distributions.  In Sec.~\ref{sec:Higgs} we discuss Higgs, top-pair and jet production, and the sensitivity to the gluon distribution.  In Sec.~\ref{sec:alphaS} we discuss the values of $\alpha_S$ obtained from DIS data.  Finally, we summarise in Sec.~\ref{sec:summary}.

\section{Status of MSTW PDF analysis}
\label{sec:mstw}

\begin{figure}[t]
  \vspace{-5cm}
  \includegraphics[width=0.5\textwidth]{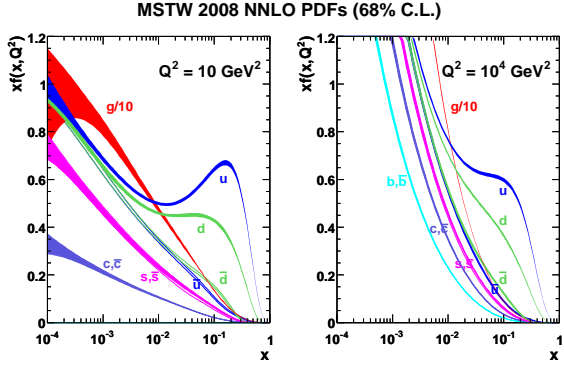}
  \caption{MSTW 2008 NNLO PDFs at two different $Q^2$ values~\cite{Martin:2009iq}.\label{fig:mstw2008}}
\end{figure}
\begin{figure}[t]
  \vspace{-5cm}
  \includegraphics[width=0.5\textwidth]{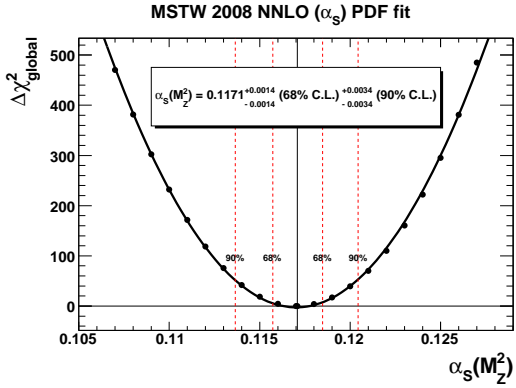}
  \caption{$\Delta\chi^2_{\rm global}$ as a function of $\alpha_S(M_Z^2)$~\cite{Martin:2009bu}.\label{fig:chisqAlphaS}}
\end{figure}
\begin{figure}[t]
  \vspace{-5cm}
  \includegraphics[width=0.5\textwidth]{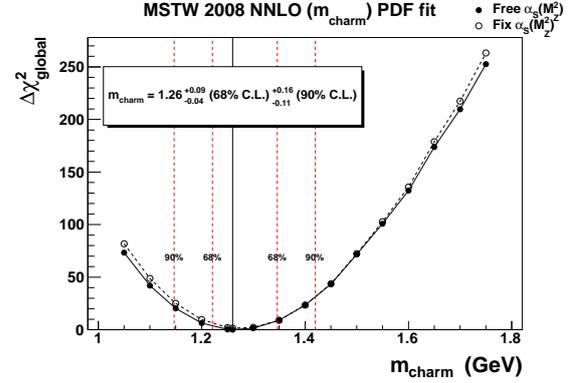}
  \caption{$\Delta\chi^2_{\rm global}$ as a function of (pole-mass) $m_c$~\cite{Martin:2010db}.\label{fig:chisqMCharm}}
\end{figure}
\begin{table}[t]
  \begin{tabular}{|l|c|c|c|}
    \hline
    LHC, $\sqrt{s} = 7$ TeV & $\sigma_W$ & $\sigma_Z$ & $\sigma_H$ \\
    \hline
    PDF-only uncertainty & $^{+1.7\%}_{-1.6\%}$ & $^{+1.7\%}_{-1.5\%}$ & $^{+1.1\%}_{-1.6\%}$ \\ \hline
    PDF+$\alpha_S$ uncertainty & $^{+2.5\%}_{-1.9\%}$ & $^{+2.5\%}_{-1.9\%}$ & $^{+3.7\%}_{-2.9\%}$ \\ \hline
    PDF+$\alpha_S$+$m_{c,b}$ uncertainty & $^{+2.7\%}_{-2.2\%}$ & $^{+2.9\%}_{-2.4\%}$ & $^{+3.7\%}_{-2.9\%}$ \\
    \hline
  \end{tabular}
  \caption{Impact of $\alpha_S$ and $m_{c,b}$ variation on LHC cross sections~\cite{Martin:2010db}.\label{tab:uncertainty}}
\end{table}

The ``MSTW 2008'' determination of PDFs~\cite{Martin:2009iq} at leading-order (LO), next-to-leading order (NLO) and next-to-next-to-leading order (NNLO) superseded the previously available ``MRST'' PDFs.  New data sets fitted included neutrino structure functions ($F_2$ and $xF_3$) from NuTeV and CHORUS, neutrino dimuon cross sections from CCFR and NuTeV, HERA data on $F_2^{\rm charm}$ and on inclusive jet production in DIS, and Tevatron Run II data on inclusive jet production, the lepton charge asymmetry from $W$ decays and the $Z$ rapidity distribution.  The CCFR/NuTeV dimuon cross sections allowed the strange-quark and -antiquark distributions to be fit directly for the first time.  The Tevatron Run II jet data were found to prefer a softer gluon distribution at high $x$ than the previous Run I data used in the MRST 2001--2006 fits.  Uncertainties on the PDFs, shown in Fig.~\ref{fig:mstw2008}, were propagated from the experimental errors on the fitted data points using a new dynamic procedure for each eigenvector of the covariance matrix.  Subsequent studies used the same procedure to determine the experimental error on the best-fit $\alpha_S(M_Z^2)$, shown in Fig.~\ref{fig:chisqAlphaS}~\cite{Martin:2009bu}, and on the best-fit (pole-mass) $m_c$, shown in Fig.~\ref{fig:chisqMCharm}~\cite{Martin:2010db}.

Uncertainties in both $\alpha_S(M_Z^2)$ and the heavy-quark masses $m_{c,b}$ induce an additional uncertainty in cross-section predictions compared to the uncertainty arising only from PDFs.  This increase in uncertainty is shown in Table~\ref{tab:uncertainty} for the $W$, $Z$ and $gg\to H$ ($M_H=120$ GeV) NNLO total cross sections at the LHC with $\sqrt{s}=7~{\rm TeV}$.  Here, the PDF uncertainties are at 68\% confidence-level (C.L.), and the parameters are varied in the ranges $\alpha_S(M_Z^2)=0.1171\pm0.0014$, $m_c=1.40\pm0.15$~GeV and $m_b=4.75\pm0.25$~GeV.  Varying $m_c$ leads to a change in $\sigma_{W,Z}$ of just over 1\%, while varying $m_b$ leads to a negligible change (0.1\%) in $\sigma_{W,Z}$, and $\sigma_H$ is insensitive to $m_{c,b}$ variation.  Adding the $m_{c,b}$ uncertainty in quadrature to the ``PDF+$\alpha_S$'' uncertainty does not result in a significant enhancement to the PDF+$\alpha_S$ uncertainty and it will therefore not be included in the remainder of this write-up.  Further preliminary studies~\cite{Watt:2010qt,Thorne:2010kj} investigated problems in describing the precise $W\to\ell\nu$ charge asymmetry data at the Tevatron, and examined the impact of including the combined HERA I data~\cite{HERA:2009wt} in the MSTW global fit, where the changes were not large enough to warrant an immediate update~\cite{Thorne:2010kj}.

\section{Benchmark exercise}
\label{sec:bench}

\begin{table*}[t]
  \centering
  \begin{tabular}{|l|c|c|c|c|c|c|}
    \hline
    & {\bf MSTW08} & {\bf CT10} & {\bf NNPDF2.1} & HERAPDF1.5 & ABKM09 & GJR08/JR09 \\
    \hline
        { HERA DIS} & \ding{52} & \ding{52} & \ding{52} & \ding{52} & \ding{52} & \ding{52} \\
        { Fixed-target DIS} & \ding{52} & \ding{52} & \ding{52} & \ding{55} & \ding{52} & \ding{52} \\
        { Fixed-target DY} & \ding{52} & \ding{52} & \ding{52} & \ding{55} & \ding{52} & \ding{52} \\
        { Tevatron $W$,$Z$} & \ding{52} & \ding{52} & \ding{52} & \ding{55} & \ding{55} & \ding{55} \\
        { Tevatron jets} & \ding{52} & \ding{52} & \ding{52} & \ding{55} & \ding{55} & \ding{52}/\ding{55} \\ \hline
        { GM-VFNS} & \ding{52} & \ding{52} & \ding{52} & \ding{52} & \ding{55} & \ding{55} \\
        { NNLO} & \ding{52} & \ding{55} & \ding{52} & \ding{52} & \ding{52} & \ding{52} \\
        \hline
  \end{tabular}
  \caption{Comparison of major PDF sets considered, and their gross features distinguished by the main classes of data included (upper part of table) and important aspects of the theoretical treatment (lower part of table), specifically regarding the treatment of heavy quarks in DIS and the provision of NNLO PDFs.  More refined differences between PDF sets are described in the text.\label{tab:compare}}
\end{table*}

Various fitting groups currently produce PDF sets: MSTW08~\cite{Martin:2009iq}, CTEQ6.6/CT10~\cite{Nadolsky:2008zw,Lai:2010vv}, NNPDF2.1~\cite{Ball:2011mu,Ball:2011uy}, HERAPDF1.0/1.5~\cite{HERA:2009wt,HERA:2010,HERA:2011}, ABKM09~\cite{Alekhin:2009ni}, GJR08/JR09~\cite{Gluck:2007ck,Gluck:2008gs,JimenezDelgado:2008hf,JimenezDelgado:2009tv}.  Past experience has shown that results obtained with the different PDF sets often do not agree within the quoted uncertainties.  Quantifying, understanding, then hopefully resolving differences in PDFs \emph{between} groups is therefore as important, if not more important, as continued improvements in PDFs \emph{within} groups.  Some recent work has been initiated by the activities of the \emph{LHC Higgs Cross Section Working Group} and the \emph{PDF4LHC Working Group}.  In particular, an exercise was proposed to use the most recent public NLO PDFs from all fitting groups to calculate some LHC benchmark processes at $\sqrt{s}=7$~TeV, specifically total cross sections for production of $W^\pm$, $Z^0$, $t\bar{t}$ and $gg\to H$ for $M_H=\{120,180,240\}$~GeV.  The aims were to establish the degree of compatibility and identify outliers amongst PDF sets, and to compare cross sections at the \emph{same} $\alpha_S$ values, thereby showing to what extent differences in predictions are due to the different $\alpha_S$ values adopted by each group, rather than differences in the PDFs themselves.  The results at NLO, initially presented by G.W.~in a PDF4LHC meeting at CERN on 26th March 2010, formed the basis for the subsequent PDF4LHC \emph{Interim Report}~\cite{Alekhin:2011sk} and PDF4LHC \emph{Interim Recommendations}~\cite{Botje:2011sn} used in the \emph{Handbook of LHC Higgs Cross Sections}~\cite{LHCHiggsCrossSectionWorkingGroup:2011ti}.  An update of the NLO comparisons to include the recent CT10~\cite{Lai:2010vv} and NNPDF2.1~\cite{Ball:2011mu} analyses, and an extension of all comparisons to NNLO, was subsequently made in a later publication~\cite{Watt:2011kp}.  In this write-up we will make a further update to the comparison plots to include the even more recent NNPDF2.1 NNLO~\cite{Ball:2011uy} and HERAPDF1.5 NNLO~\cite{HERA:2011} analyses, and we will compare the predictions to the latest LHC data on the total cross sections for $W$, $Z$ and $t\bar{t}$ production.

\begin{figure*}[t]
  \vspace*{-5cm}
  \includegraphics[width=0.5\textwidth]{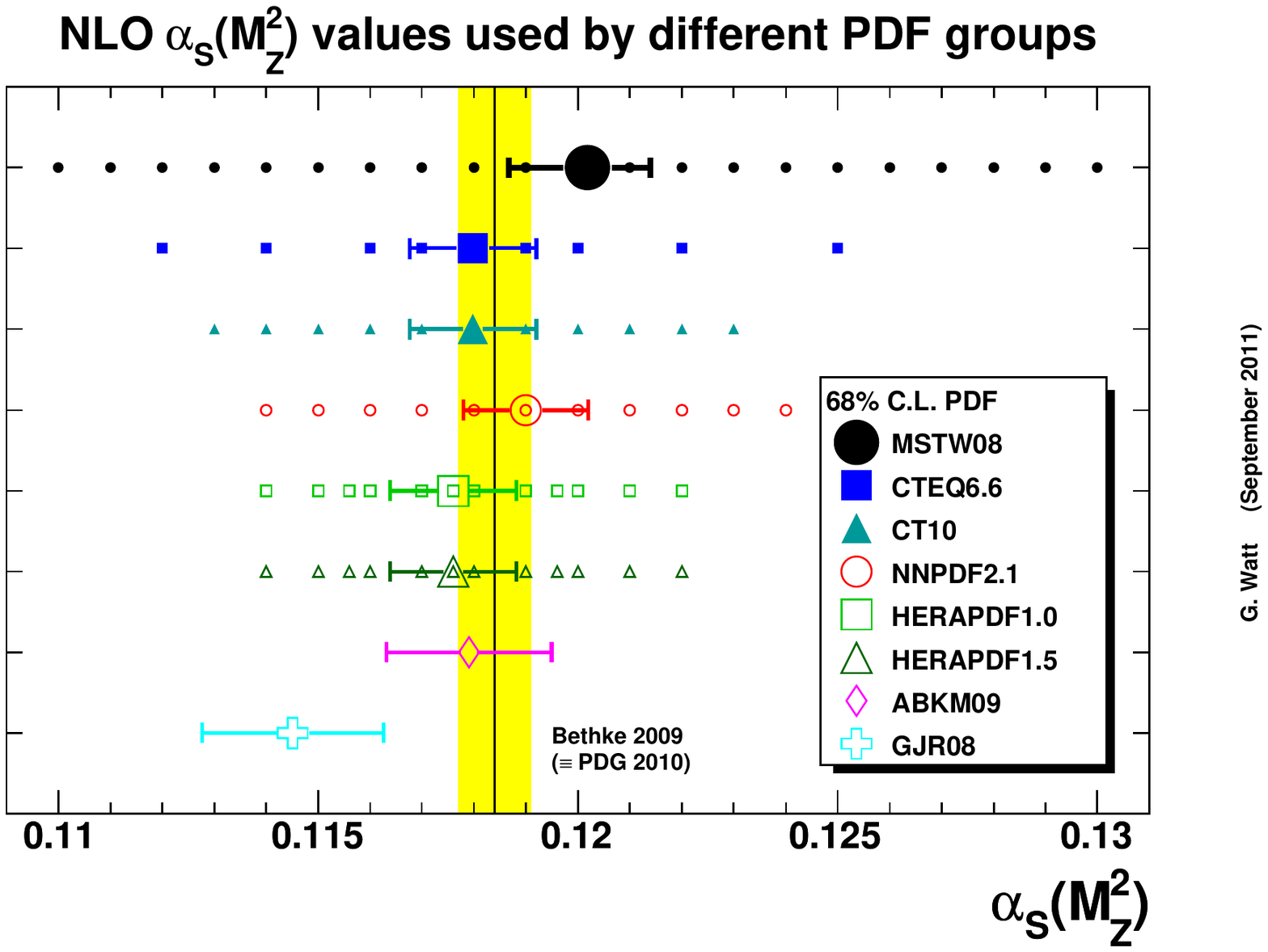}
  \includegraphics[width=0.5\textwidth]{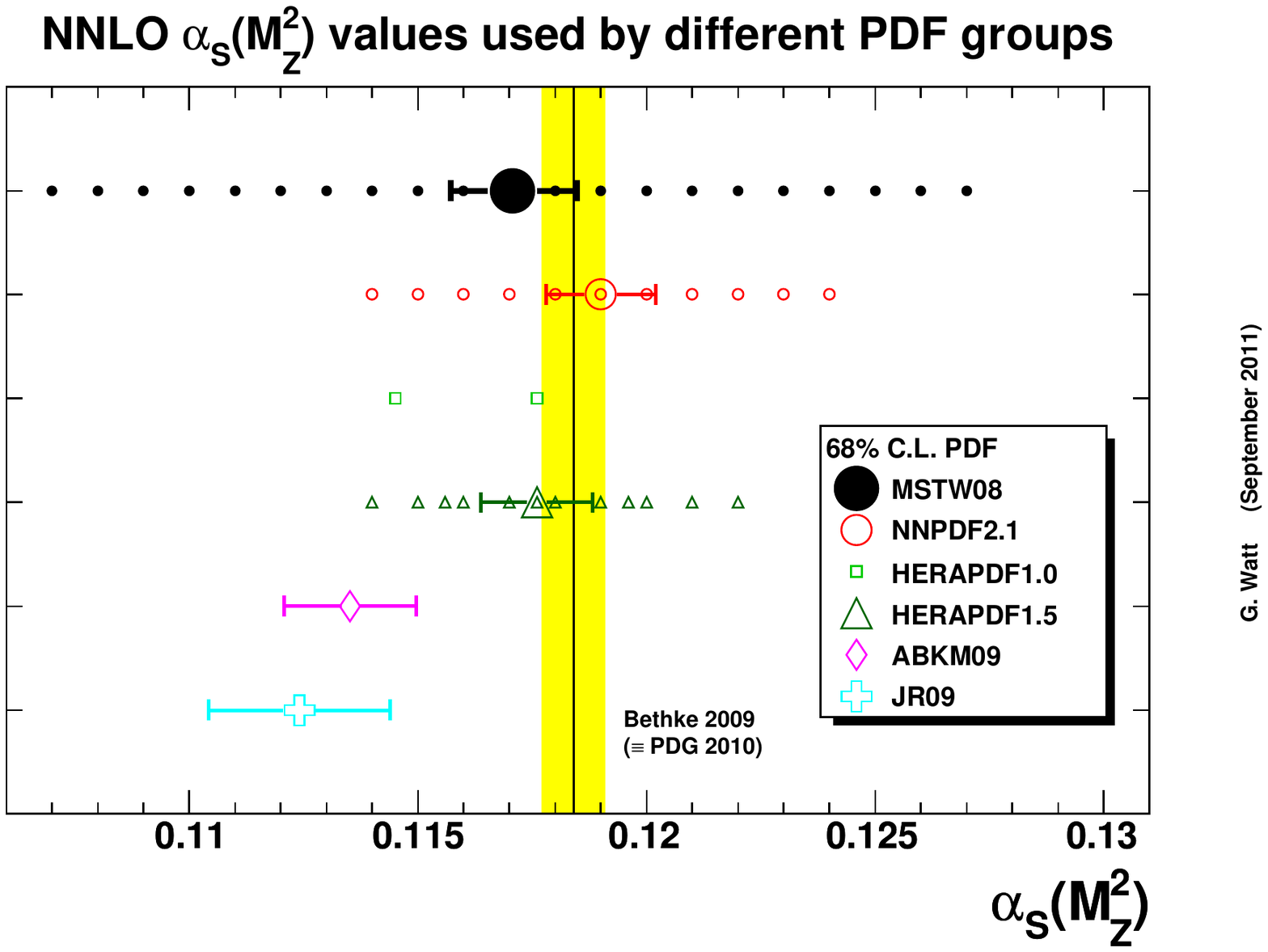}
  \caption{Values of $\alpha_S(M_Z^2)$, and their 1-$\sigma$ uncertainties, used by different PDF fitting groups at NLO and NNLO.  The smaller symbols indicate the PDF sets with alternative values of $\alpha_S(M_Z^2)$ provided by each fitting group.  The shaded band indicates the Bethke 2009 world average $\alpha_S(M_Z^2)$.\label{fig:alphaSMZ}}
\end{figure*}

We will consider only \emph{public} sets, defined to be those available for use with the latest \textsc{lhapdf} V5.8.6~\cite{Whalley:2005nh}.  The broad distinctions between data sets fitted and aspects of the theoretical treatment are summarised in Table~\ref{tab:compare}.  Only three groups (MSTW, CT and NNPDF) make fully \emph{global} fits to HERA and fixed-target DIS data, fixed-target Drell--Yan production, and Tevatron data on $W$, $Z$ and jet production, although GJR08 includes all these processes other than Tevatron $W$ and $Z$ production.  The HERAPDF1.0 fit includes \emph{only} the combined HERA I inclusive data~\cite{HERA:2009wt}, while the HERAPDF1.5 fit additionally includes the preliminary combined HERA II inclusive data.  The CT10 and NNPDF2.1 global fits include the combined HERA I inclusive data, while the other fits (MSTW08, CTEQ6.6, ABKM09, GJR08/JR09) include the older separate data from H1 and ZEUS.  The MSTW08, CT10, NNPDF2.1 and GJR08 fits include Tevatron Run II data, while CTEQ6.6 uses only Tevatron Run I data.  The original NNPDF2.1 fit has been reweighted to include Tevatron and LHC data on the $W\to\ell\nu$ charge asymmetry, denoted NNPDF2.2~\cite{Ball:2011gg}, but this reweighted PDF set will not be considered here.  Most groups now treat the heavy-quark contribution to DIS structure functions using a general-mass variable flavour number scheme (GM-VFNS), other than ABKM09 and GJR08/JR09 who use a fixed flavour number scheme (FFNS).  The change from the inadequate zero-mass variable flavour number scheme (ZM-VFNS) to the GM-VFNS was the major improvement between NNPDF2.0~\cite{Ball:2010de} and NNPDF2.1~\cite{Ball:2011mu}, now allowing a meaningful comparison to other NLO global fits.  The NNPDF fits parameterise the starting distributions at $Q_0^2=2$~GeV$^2$ as neural networks and use Monte Carlo methods for experimental error propagation.  The other groups all use the more traditional approach of parameterising the input PDFs as some functional form in $x$, each with a handful of free parameters, and use the Hessian method for experimental error propagation with differing values of the tolerance $\Delta\chi^2$, that is, the change in the goodness-of-fit measure relative to the best-fit value.  Contrary to the ``standard'' input parameterisation at $Q_0^2\ge1$~GeV$^2$, the GJR08/JR09 sets use a ``dynamical'' input parameterisation of valence-like input distributions at an optimally chosen $Q_0^2<1$~GeV$^2$, which gives a slightly worse fit quality and lower $\alpha_S$ values than the corresponding ``standard'' parameterisation, but is nevertheless favoured by the GJR08/JR09 authors.  Public NNLO fits are available from MSTW08, NNPDF2.1, HERAPDF1.0/1.5, ABKM09 and JR09.  (The first NNLO fits from the CT group should be available soon.)  The Tevatron jet cross sections are excluded from the JR09 fit, where complete NNLO corrections are unavailable, whereas they are included in the MSTW08 and NNPDF2.1 NNLO fits by making the approximation of using the NLO partonic cross section supplemented by 2-loop threshold corrections~\cite{Kidonakis:2000gi}.

In Fig.~\ref{fig:alphaSMZ} we show the default values of $\alpha_S(M_Z^2)$ used by different fitting groups at NLO and NNLO, and we compare to the world average value obtained by S.~Bethke in 2009~\cite{Bethke:2009jm} (and taken over by the Particle Data Group in 2010~\cite{Nakamura:2010zzi}).  The values for MSTW08, ABKM09 and GJR08/JR09 are obtained from a simultaneous fit with the PDF parameters, while $\alpha_S(M_Z^2)$ for other groups is applied as an external constraint, generally chosen to be close to the world average~\cite{Bethke:2009jm}, and for those groups we assume a 1-$\sigma$ uncertainty of $\pm0.0012$~\cite{Alekhin:2011sk}.  The smaller symbols indicate the PDF sets with alternative values of $\alpha_S(M_Z^2)$ values provided by each fitting group.  The fitted NLO $\alpha_S(M_Z^2)$ values are always larger than the fitted NNLO $\alpha_S(M_Z^2)$ values in an attempt by the fit to mimic the missing higher-order corrections, which are generally positive.  This trend is repeated in the recent NNPDF determinations of $\alpha_S(M_Z^2)=0.1191\pm0.0006$ at NLO~\cite{Lionetti:2011pw} and $\alpha_S(M_Z^2)=0.1173\pm0.0007$ at NNLO~\cite{Ball:2011us}, where the experimental uncertainties are obtained using $\Delta\chi^2=1$, but these are not used as the default $\alpha_S(M_Z^2)$ values for the provided NNPDF2.1 PDFs.

\section{$W$ and $Z$ production at the LHC}
\label{sec:WandZ}

\begin{figure}[t]
  \vspace*{-5cm}
  \includegraphics[width=0.5\textwidth]{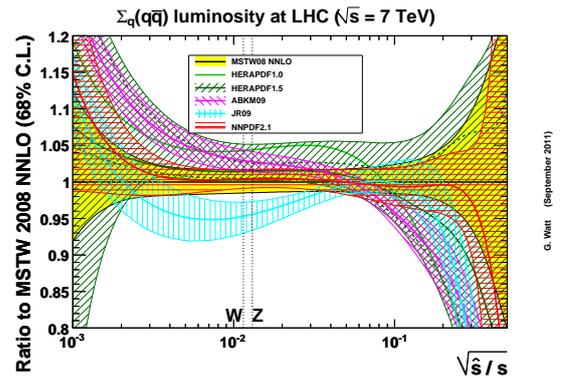}
  \caption{NNLO $q\bar{q}$ luminosities as the ratio to MSTW 2008.\label{fig:qqbarlumi}}
\end{figure}
\begin{figure*}[t]
  \vspace*{-5cm}
  \includegraphics[width=0.5\textwidth]{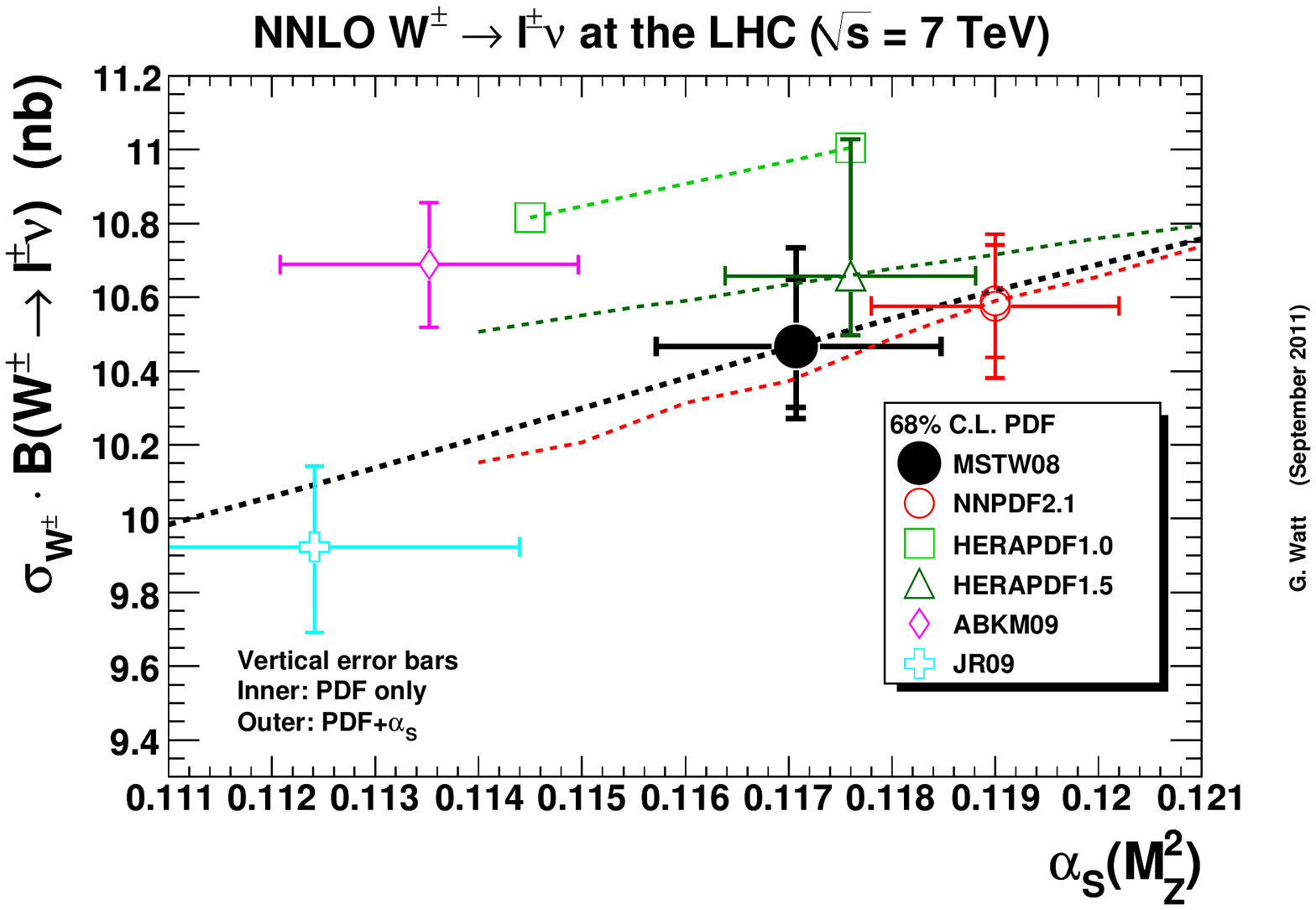}
  \includegraphics[width=0.5\textwidth]{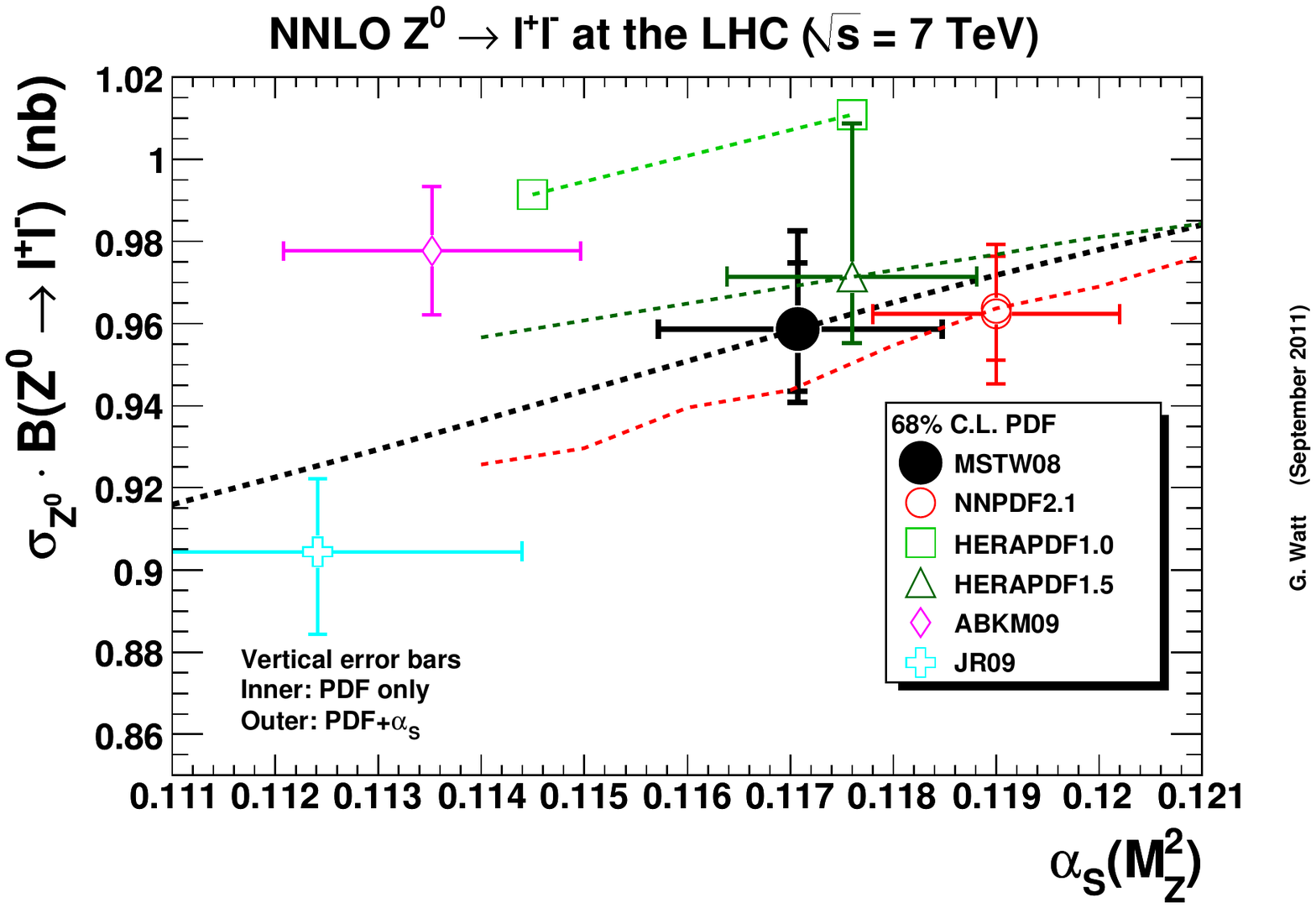}
  \caption{NNLO $W^\pm$ ($=W^++W^-$) and $Z^0$ total cross sections at the LHC, plotted as a function of $\alpha_S(M_Z^2)$.\label{fig:wzvsasmz}}
\end{figure*}

To understand properties of hadronic cross sections, such as PDF uncertainties or the dependence on collider energy, it is useful to consider the relevant parton--parton luminosities.  We define a $q\bar{q}$ luminosity, relevant for the (rapidity-integrated) total cross sections for $W$ and $Z$ production at the LHC, as
\[
\hspace*{-7mm}\frac{\partial {\cal L}_{\Sigma_q (q\bar{q})}}{\partial \hat{s}} = \frac{1}{s} \int_\tau^1\frac{{\rm d}x}{x}\hspace*{-1mm}\sum_{q=d,u,s,c,b}\hspace*{-3mm}\left[q(x,\hat{s})\bar{q}(\tau/x,\hat{s}) + (q\leftrightarrow\bar{q})\right],
\]
where $\tau\equiv\hat{s}/s$ and $\sqrt{\hat{s}}$ is the partonic centre-of-mass energy.  Note that this generic $q\bar{q}$ luminosity does not specifically include the correct flavour combinations for $W^\pm$ or $Z^0$ production.  More detailed studies would, for example, include the correct couplings of the vector bosons to quarks and antiquarks, or consider the specific $u\bar{d}$ (for $W^+$) or $d\bar{u}$ (for $W^-$) partonic luminosities.

In Fig.~\ref{fig:qqbarlumi} we show the NNLO $q\bar{q}$ luminosities as the ratio with respect to the MSTW 2008 NNLO luminosities, for the LHC at $\sqrt{s} = 7$~TeV.  We use the default $\alpha_S$ values for each set, shown in Fig.~\ref{fig:alphaSMZ}.  The HERAPDF1.0 NNLO curve (without uncertainties) is for $\alpha_S(M_Z^2)=0.1176$.  The inner uncertainty bands (dashed lines) for HERAPDF1.5 correspond to the (asymmetric) experimental errors, while the outer uncertainty bands (shaded regions) also include the model and parameterisation errors, including uncertainties on heavy-quark masses but not on $\alpha_S$.  It is not possible to separate the ``PDF only'' uncertainty for ABKM09 and JR09, therefore the uncertainty bands for those sets also include the $\alpha_S$ uncertainty, and the uncertainty bands for ABKM09 also include uncertainties on heavy-quark masses.  This is undesirable but unavoidable given that these groups do not provide PDF sets for fixed $\alpha_S$ (and fixed $m_{c,b}$ for ABKM09).  The relevant values of $\sqrt{\hat{s}} = M_{W,Z}$ are indicated, and there is good agreement for the two global fits (MSTW08 and NNPDF2.1), but more variation for the other sets.  The NNLO trend between groups is similar to at NLO~\cite{Watt:2011kp,mstwpdf}, with the exception of HERAPDF at large $\hat{s}$ values, where the HERAPDF NLO sets have a much larger $q\bar{q}$ luminosity than other NLO PDF groups.

In Fig.~\ref{fig:wzvsasmz} we show the $W^\pm$ ($=W^++W^-$) and $Z^0$ total cross section multiplied by the appropriate leptonic branching ratios, $B(W^\pm\to\ell^\pm\nu)$ or $B(Z^0\to\ell^+\ell^-)$, calculated at NNLO~\cite{Hamberg:1990np} with a scale choice $\mu_R=\mu_F=M_{W,Z}$, plotted as a function of $\alpha_S(M_Z^2)$.  The markers are centred on the default $\alpha_S(M_Z^2)$ value and the corresponding predicted cross section of each PDF fitting group.  The horizontal error bars span the $\alpha_S(M_Z^2)$ uncertainty, the inner vertical error bars span the ``PDF only'' uncertainty where possible (i.e.~not for ABKM09 or JR09), and the outer vertical error bars span the ``PDF+$\alpha_S$'' uncertainty.  The effect of the additional $\alpha_S$ uncertainty is small for $W$ and $Z$ production.  The dashed lines interpolate the cross-section predictions calculated with the alternative PDF sets with different $\alpha_S(M_Z^2)$ values provided by each group, represented by the smaller symbols in Fig.~\ref{fig:alphaSMZ}.  The two global fits (MSTW08 and NNPDF2.1) are in good agreement, as was apparent from the $q\bar{q}$ luminosity plot in Fig.~\ref{fig:qqbarlumi}.  The central HERAPDF1.5 prediction is close to the global fits at NNLO, contrary to the predictions using HERAPDF1.0 NNLO or HERAPDF1.0/1.5 NLO~\cite{Watt:2011kp,mstwpdf}.

\begin{figure*}[t]
  \vspace*{-3cm}
  \includegraphics[width=0.5\textwidth]{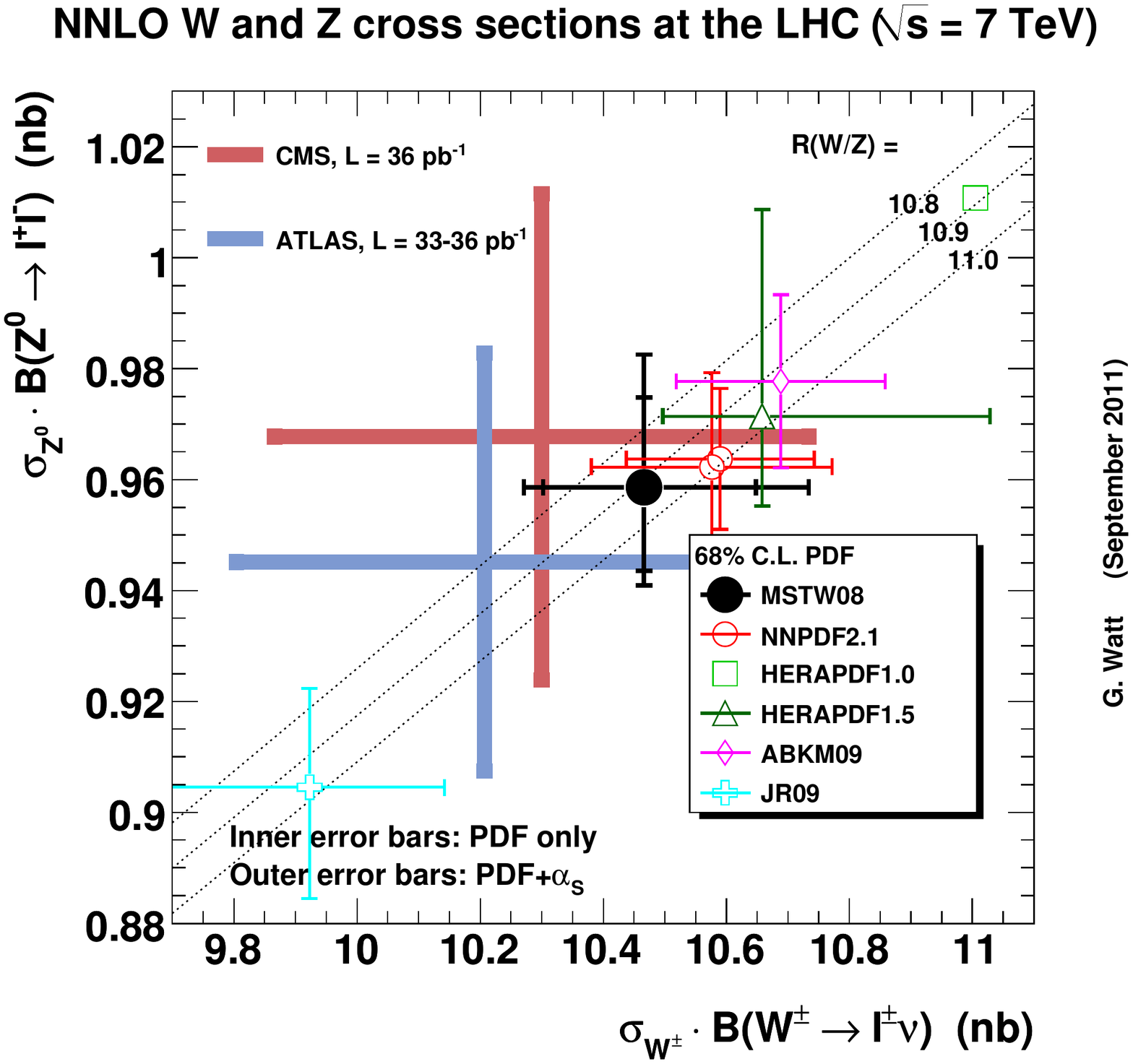}
  \includegraphics[width=0.5\textwidth]{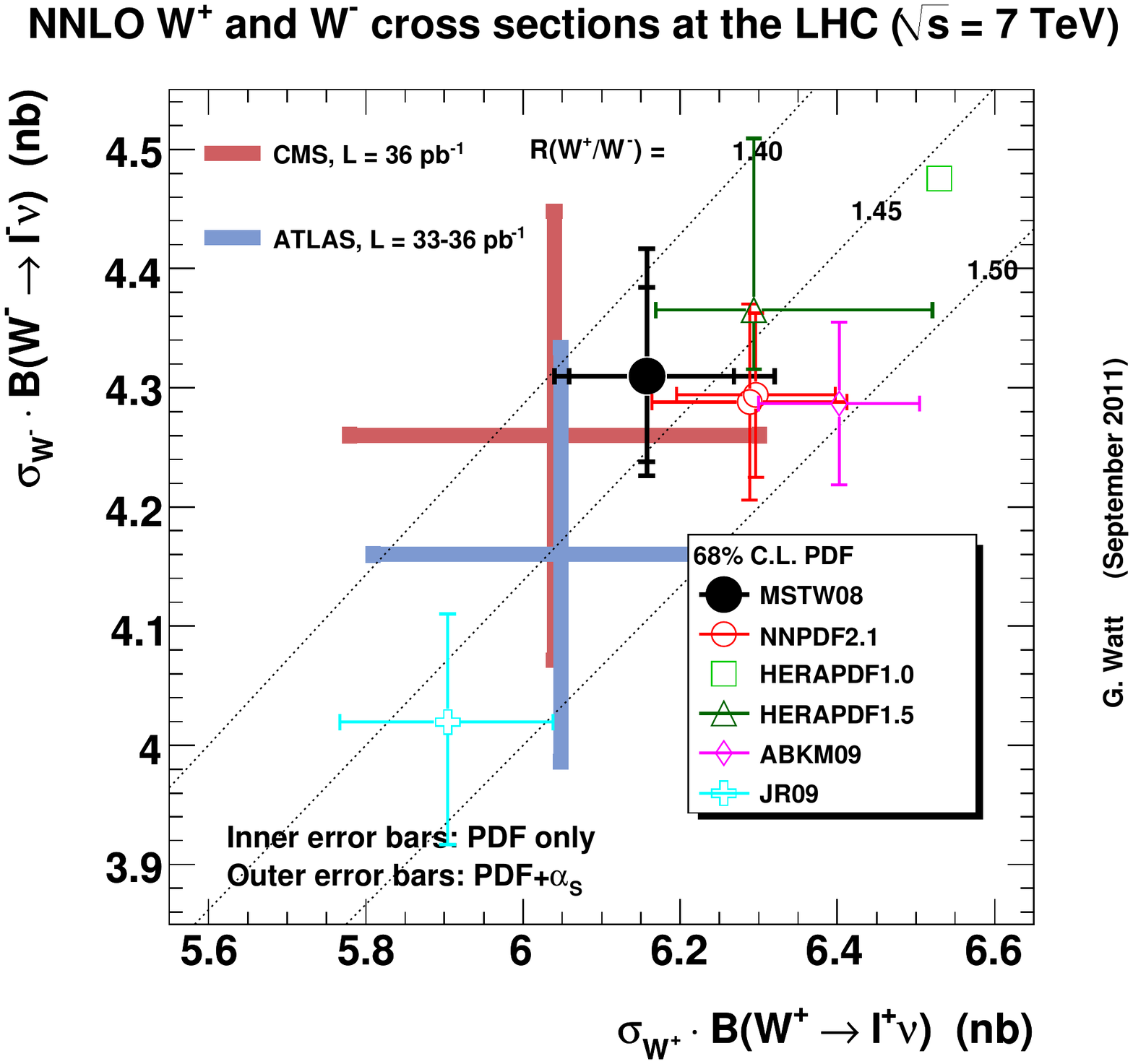}
  \caption{$W^\pm$ vs.~$Z^0$ and $W^+$ vs.~$W^-$ total cross sections at NNLO, compared to data from CMS~\cite{Chatrchyan:2011nx} and ATLAS~\cite{Aad:2011dm}.\label{fig:wzerr}}
\end{figure*}
\begin{figure*}[t]
  \vspace*{-3cm}
  \includegraphics[width=0.5\textwidth]{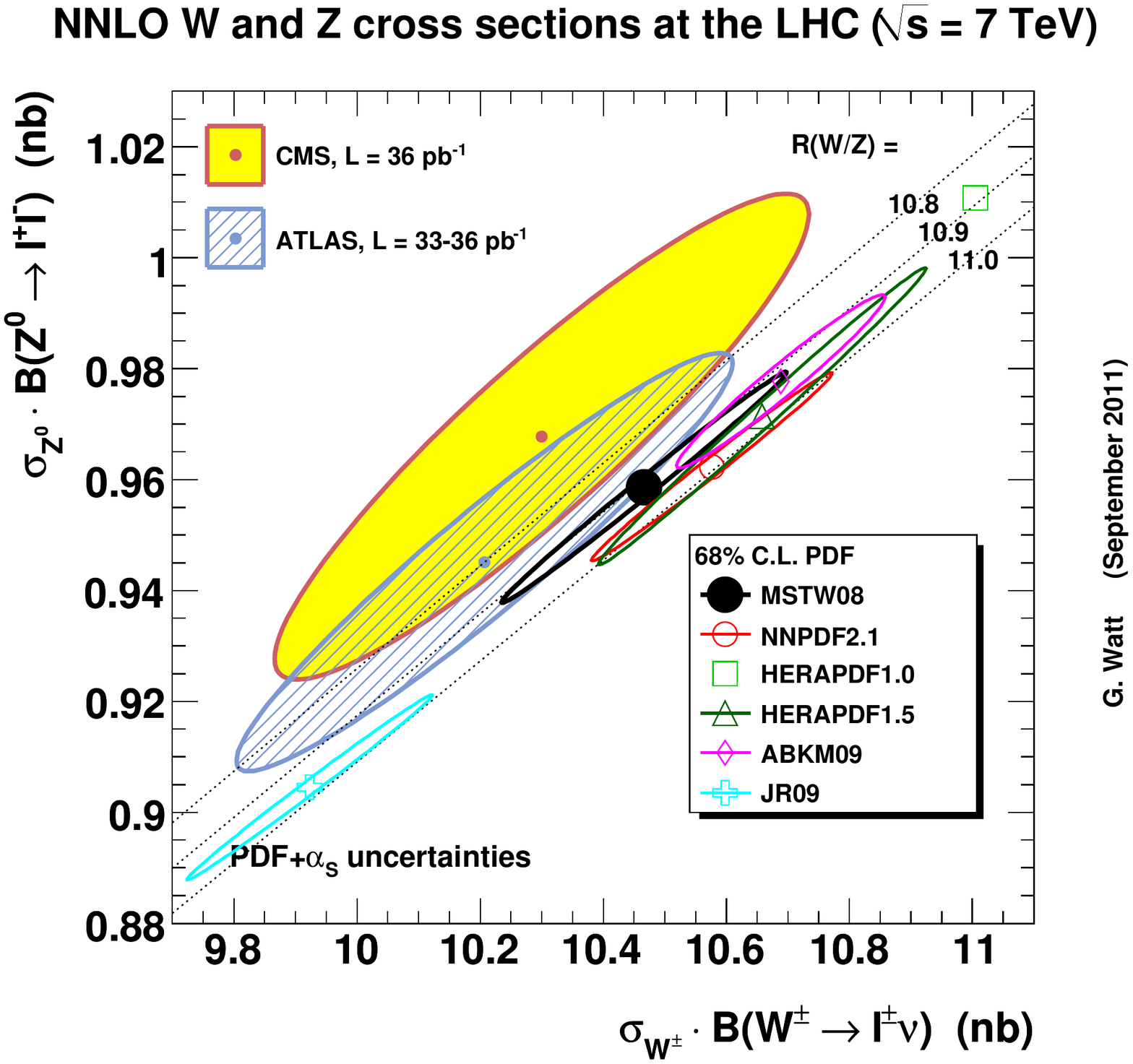}
  \includegraphics[width=0.5\textwidth]{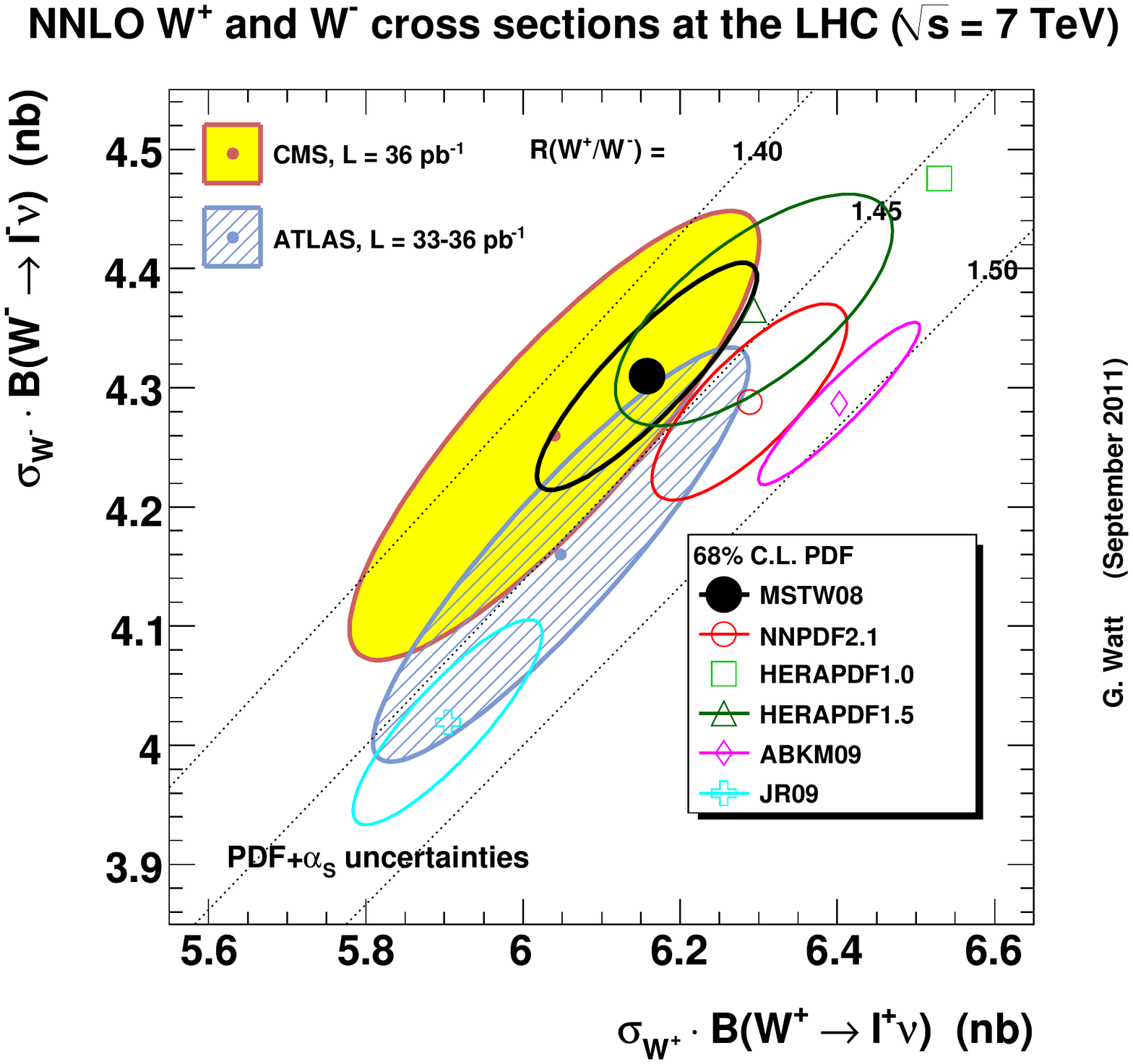}
  \caption{Same as Fig.~\ref{fig:wzerr}, but with ellipses accounting for the correlations between the two cross sections.\label{fig:wzcon}}
\end{figure*}

The total cross-section \emph{ratios}, $W^\pm/Z^0$ and $W^+/W^-$, are insensitive to NNLO corrections and the value of $\alpha_S(M_Z^2)$.  The $W^\pm$ ($=W^++W^-$) and $Z^0$ total cross sections are highly correlated, which can be understood by considering the dominant partonic contributions arising from $u$ and $d$ quarks, i.e.
\begin{equation}
  \label{eq:WoverZ}
  \frac{\sigma_{W^+}+\sigma_{W^-}}{\sigma_{Z^0}}\sim \frac{u(x_1)+d(x_1)}{0.29\,u(\tilde{x}_1)+0.37\,d(\tilde{x}_1)},
\end{equation}
where we have neglected the contributions with $q\leftrightarrow\bar{q}$, assuming that $q(x_1)\bar{q}(x_2)$ dominates over $\bar{q}(x_1)q(x_2)$, and the numerical values in the denominator are the appropriate sums of the squares of the vector and axial-vector couplings.  We have also assumed that $\bar{u}(x_2)\approx\bar{d}(x_2)$.  Here, the momentum fractions are $x_{1,2}=(M_W/\sqrt{s})\exp(\pm y_W)$ and $\tilde{x}_{1,2}=(M_Z/\sqrt{s})\exp(\pm y_Z)$, and $y_W$ or $y_Z$ should be interpreted as some ``average'' rapidity appropriate for the (rapidity-integrated) total cross section.  The combination of $u$- and $d$-quark contributions is very similar (in numerical prefactors, in $x$ values, and in $Q^2=M_{W,Z}^2$ values) in both the numerator and denominator of Eq.~(\ref{eq:WoverZ}), therefore the PDF dependence almost cancels in the $W^\pm/Z^0$ ratio.  The $W^+$ and $W^-$ total cross sections are much less correlated, since
\[
  \frac{\sigma_{W^+}}{\sigma_{W^-}}\sim\frac{u(x_1)\,\bar{d}(x_2)}{d(x_1)\,\bar{u}(x_2)}\sim\frac{u(x_1)}{d(x_1)},
\]
and therefore the $W^+/W^-$ cross-section ratio is a sensitive probe of the $u/d$ ratio and there is more variation between different PDF sets.

In Fig.~\ref{fig:wzerr} we show $W^\pm$ ($=W^++W^-$) versus $Z^0$ and $W^+$ versus $W^-$ total cross sections, and we draw dotted lines where the ratio of cross sections, $W^\pm/Z^0$ or $W^+/W^-$, is constant.  We also compare to the experimental measurements using the 2010 LHC data from CMS~\cite{Chatrchyan:2011nx} and ATLAS~\cite{Aad:2011dm}.  Here, the measured $Z^0$ cross sections have been corrected~\cite{Watt:2011kp} for the small $\gamma^*$ contribution and the finite invariant-mass range of the lepton pair (different for ATLAS and CMS) using a theory calculation at NNLO~\cite{Anastasiou:2003ds}.  The spread in predictions using the different PDF sets is comparable to the (dominant) luminosity uncertainty of 4\% (CMS) or 3.4\% (ATLAS), and perhaps the two extreme predictions (HERAPDF1.0 and JR09) are somewhat disfavoured.  In Fig.~\ref{fig:wzcon} we show the same plots with ellipses drawn to account for the correlations between the two cross sections, both for the experimental measurements and for the theoretical predictions.  Here, the ellipses are defined such that the projection onto either axes gives the 1-$\sigma$ uncertainty for the individual cross sections, meaning that the area of the two-dimensional ellipse corresponds to a confidence-level somewhat smaller than the conventional 68\%.  The largest uncertainty in the ATLAS and CMS total cross-section ratios comes from extrapolating the measured (fiducial) cross section over the whole phase space.  Indeed, improvements in the acceptance calculation led to the central value of the ATLAS $W^+/W^-$ total cross-section ratio shifting from the preliminary result of 1.51 in March 2011~\cite{ATLAS:WandZ} to 1.45 in the final publication~\cite{Aad:2011dm} following observations made in Ref.~\cite{Watt:2011kp}.  Therefore, data-to-theory comparisons are best made at the level of the fiducial cross section (i.e.~within the acceptance), which is now possible using the public \textsc{fewz}~\cite{Melnikov:2006kv,Gavin:2010az} and \textsc{dynnlo}~\cite{Catani:2009sm} codes, and indeed was done in the ATLAS publication~\cite{Aad:2011dm}.

Of course, as a constraint on PDFs, differential distributions are more useful than total or fiducial cross sections.  The $W^\pm$ charge asymmetry is sensitive to the difference between up- and down-valence quark distributions, i.e.
\begin{eqnarray*}
  A_W(y_W) &=& \frac{{\rm d}\sigma(W^+)/{\rm d}y_W-{\rm d}\sigma(W^-)/{\rm d}y_W}{{\rm d}\sigma(W^+)/{\rm d}y_W+{\rm d}\sigma(W^-)/{\rm d}y_W} \\ &\approx& \frac{u_v(x_1)-d_v(x_1)}{u(x_1)+d(x_1)}.
\end{eqnarray*}
Experimentally, it is not possible to directly reconstruct the $W$ rapidity since the longitudinal momentum of the decay neutrino is unknown, so instead the $W^\pm\to\ell^\pm\nu$ charge asymmetry is measured as a function of the charged-lepton pseudorapidity $\eta_\ell$, i.e.
\begin{eqnarray*}
  A_\ell(\eta_\ell) &=& \frac{{\rm d}\sigma(\ell^+)/{\rm d}\eta_\ell-{\rm d}\sigma(\ell^-)/{\rm d}\eta_\ell}{{\rm d}\sigma(\ell^+)/{\rm d}\eta_\ell+{\rm d}\sigma(\ell^-)/{\rm d}\eta_\ell} \\ &\equiv& A_W(y_W)\otimes (W^\pm\to\ell^\pm\nu),
\end{eqnarray*}
where the last line is symbolic and indicates the underlying $W^\pm$ charge asymmetry convoluted with the $W^\pm\to\ell^\pm\nu$ decay.  Measurements of $A_\ell(\eta_\ell)$ have provided the first useful PDF constraint from LHC data, and indicate that $A_\ell(0)$ is underestimated by the MSTW08 fit and is better described by some other PDF sets, implying that $u_v-d_v$ is too small at $x\sim M_W/\sqrt{s}\sim 0.01$.  This behaviour may be traced to the independent small-$x$ powers, $xu_v\propto x^{0.29\pm0.02}$ and $xd_v\propto x^{0.97\pm0.11}$, at the input scale $Q_0^2=1$~GeV$^2$ for the MSTW08 NLO fit, compared to some other groups (e.g.~CTEQ6.6/CT10) where the small-$x$ powers of $u_v$ and $d_v$ are assumed to be equal.  On the other hand, this implies some tension between the LHC $W\to\ell\nu$ charge asymmetry data and the data already included in the MSTW08 fit (e.g.~the Tevatron $W\to\ell\nu$ asymmetry, the NMC $F_2^d/F_2^p$ ratio, and the E866/NuSea Drell--Yan $\sigma^{pd}/\sigma^{pp}$ ratio).  Other tensions have been observed with the precise Tevatron data on the $W\to\ell\nu$ charge asymmetry, and partially resolved by more flexible nuclear corrections for deuteron structure functions~\cite{Thorne:2010kj}.  Further attempts to resolve these tensions will be necessary for any future update of the MSTW08 fit.  The ATLAS Collaboration provide differential cross sections for $W^+\to\ell^+\nu$ and $W^-\to\ell^-\bar{\nu}$ with the complete information on correlated systematic uncertainties, which is potentially more useful for PDF fits than simply $A_\ell(\eta_\ell)$, and the individual $W^+\to\ell^+\nu$ and $W^-\to\ell^-\bar{\nu}$ cross sections seem to be better described by the MSTW08 fit than many other PDF sets~\cite{Aad:2011dm}.

Currently, perhaps the least well-known parton distributions are the strange-quark and -antiquark distributions, where the only experimental information comes from the CCFR/NuTeV dimuon cross sections in the region $0.01\lesssim x\lesssim 0.2$.  Although these data are included in the MSTW08, CT10 and NNPDF2.1 analysis, the three groups obtain quite different results for the $s$ and $\bar{s}$ distributions in the data region.  These differences are not well understood, but may be due to issues such as the acceptance calculation, different treatment of nuclear corrections, or differences in the treatment of charm production in charged-current DIS.  In particular, the NNPDF2.1 fit includes the contribution initiated by the charm-quark PDF, leading to a smaller strange-quark PDF compared to the MSTW08 and CT10 fits, which do not include this contribution.  In the $x$ region outside of the CCFR/NuTeV data, the $s$ and $\bar{s}$ distributions are largely determined by the assumed parameterisation (with the possible exception of NNPDF).  A complementary, and perhaps theoretically cleaner, process to probe strangeness at the LHC is $W$ production with an associated charm-tagged jet.  The dominant partonic subprocesses are $\bar{s}\,g\to W^+\,\bar{c}$ and $s\,g\to W^-\,c$, with a small Cabibbo-suppressed contribution from $\bar{d}\,g\to W^+\,\bar{c}$ (5\%) and $d\,g\to W^-\,c$ (15\%).  A first preliminary measurement has been made by CMS~\cite{CMS:Wcharm} of the cross-section ratios $R_c\equiv\sigma(W+c)/\sigma(W+{\rm jets})$, probing the strange content of the proton relative to other light-quark flavours, and $R_c^\pm\equiv\sigma(W^++\bar{c})/\sigma(W^-+c)$, potentially probing the strange asymmetry.  The current CMS data~\cite{CMS:Wcharm} do not strongly discriminate between current PDF sets, although the measurement of $R_c$ is in slightly better agreement with MSTW08 and CT10 and is more than 1-$\sigma$ above the NNPDF2.1 prediction.  With more precise measurements to come, including differential distributions, the $W$+charm process should enable powerful constraints to be made on the $s$ and $\bar{s}$ distributions.

\section{Higgs, top-pair and jet production}
\label{sec:Higgs}

Whereas the cross sections for production of $W$ and $Z$ bosons are sensitive to the \emph{quark} distributions, we now turn to processes that are sensitive to the \emph{gluon} distribution.  Of course, one of the main goals of the initial LHC physics programme is to either discover or exclude the Standard Model (SM) Higgs boson ($H$).  This requires precise knowledge of the theoretical cross section, where the dominant production mechanism at both the Tevatron and LHC is gluon--gluon fusion ($gg\to H$) through a top-quark loop, with the gluon distribution being sampled at $x\sim M_H/\sqrt{s}$.  The exclusion limits at 95\% C.L.~from December 2011~\cite{TEVNPH:2011cb,ATLAS:Higgs,CMS:Higgs} are given in Table~\ref{tab:MHrange}, together with the relevant $x$ values probed.
\begin{table}[b]
  \centering
  \begin{tabular}{|c|c|c|}
    \hline
    & $M_H$~(GeV) & $x\sim M_H/\sqrt{s}$ \\ \hline
    Tevatron & 156 -- 177 & 0.08 -- 0.09 \\ \hline
    & 112.7 -- 115.5 & 0.02 -- 0.02 \\
    ATLAS & 131 -- 237 & 0.02 -- 0.03 \\
    & 251 -- 453 & 0.04 -- 0.06 \\ \hline
    CMS & 127 -- 600 & 0.02 -- 0.09 \\
    \hline
  \end{tabular}
  \caption{Exclusion limits at 95\% C.L.~for the SM Higgs boson (as of December 2011)~\cite{TEVNPH:2011cb,ATLAS:Higgs,CMS:Higgs} and the approximate $x$ values probed.\label{tab:MHrange}}
\end{table}

We will calculate the total cross section ($\sigma_H$) for SM Higgs boson production (without decay) from gluon--gluon fusion via a top-quark loop, with a fixed scale choice of $\mu_R=\mu_F=M_H$.  The $m_t$ dependence is retained only at LO, with $m_t=171.3$~GeV (PDG 2009 best value), and the higher-order corrections are calculated in the limit of an infinite top-quark mass, with NNLO corrections from Ref.~\cite{Harlander:2002wh}.  We do not include the small bottom-quark loop contributions to the $gg\to H$ cross section.  The size of the higher-order corrections to the $gg\to H$ total cross sections is substantial.  Taking the appropriate MSTW08 PDFs and $\alpha_S$ values consistently at each perturbative order for $\sigma_H$ with $M_H=160$~GeV, then the NLO/LO ratio is 2.1 (Tevatron) or 1.9 (LHC), the NNLO/LO ratio is 2.7 (Tevatron) or 2.4 (LHC), and so the NNLO/NLO ratio is 1.3 (Tevatron and LHC).  The perturbative series is therefore slowly convergent, mandating the use of (at least) NNLO calculations together with the corresponding NNLO PDFs and $\alpha_S$ values.  The convergence can be improved by using a scale choice $\mu_R=\mu_F=M_H/2$, which mimics the effect of soft-gluon resummation.  However, we aim to study only the PDF and $\alpha_S$ dependence of the $gg\to H$ cross sections, and we do \emph{not} aim to come up with a single ``best'' prediction together with a complete evaluation of all sources of theoretical uncertainty.

\begin{figure*}[t]
  \vspace*{-5cm}
  \includegraphics[width=0.5\textwidth]{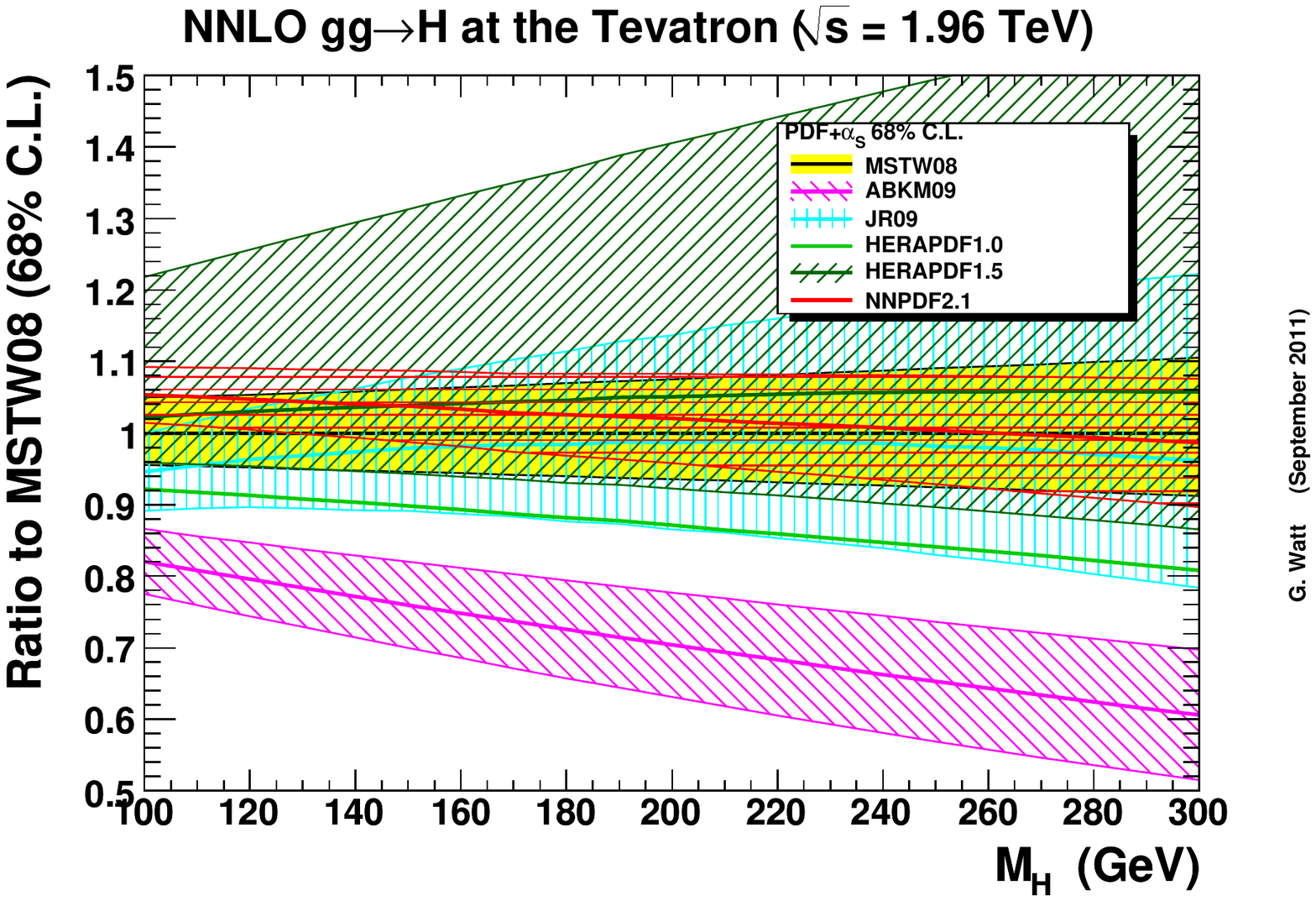}
  \includegraphics[width=0.5\textwidth]{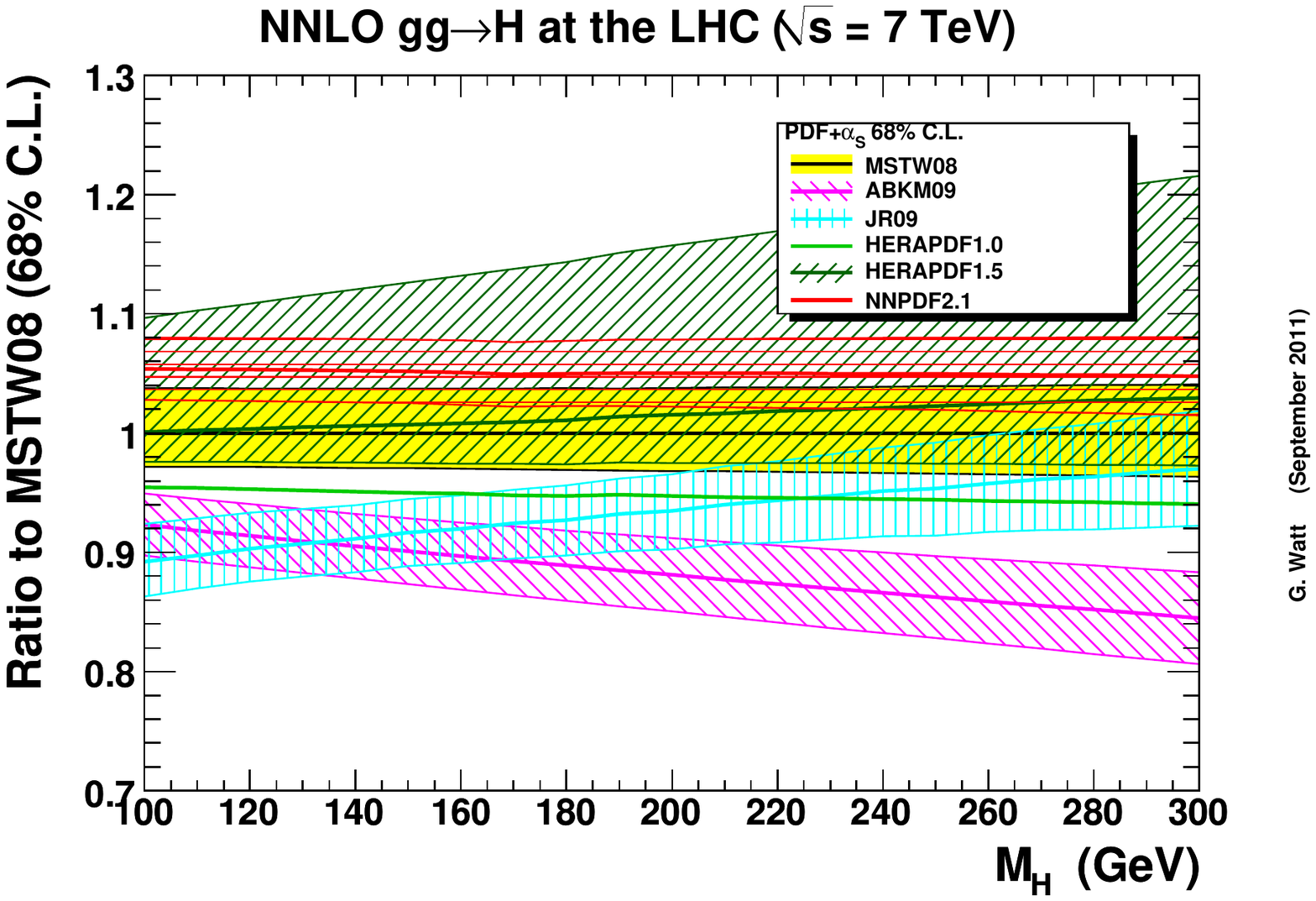}
  \caption{Ratio to MSTW 2008 NNLO $gg\to H$ total cross section, plotted as a function of $M_H$, with PDF+$\alpha_S$ uncertainty bands at 68\% C.L.\label{fig:gghvsMH}}
\end{figure*}

The ratios of the NNLO $gg\to H$ cross sections with respect to the MSTW08 predictions, plotted against the SM Higgs mass $M_H$, are shown for the Tevatron and LHC in Fig.~\ref{fig:gghvsMH}, where PDF+$\alpha_S$ uncertainty bands at 68\% C.L.~are plotted.  It can be seen that there is good agreement for the two global fits (MSTW08 and NNPDF2.1) at NNLO.  The central value of HERAPDF1.5 is also in good agreement, but it has a very large uncertainty in the upwards direction, and we will return to this feature later.  The HERAPDF1.0 prediction with $\alpha_S(M_Z^2) = 0.1176$ lies somewhat below MSTW08, NNPDF2.1 and HERAPDF1.5.  However, the ABKM09 prediction lies even further below MSTW08 at the LHC, and especially at the Tevatron, even allowing for the PDF+$\alpha_S$ uncertainties, by an amount much larger than the scale uncertainty ($\sim$10\%).  The Tevatron and LHC exclusion bounds are based on predictions using the MSTW08 PDFs, and the decision not to use \emph{all} available NNLO PDFs has drawn some misguided criticism from certain quarters~\cite{Baglio:2011wn,Alekhin:2011ey}, particularly before the LHC exclusion results became available due to the more significant discrepancies between PDF sets for Tevatron predictions.

\begin{figure*}[t]
  \vspace*{-5cm}
  \includegraphics[width=0.5\textwidth]{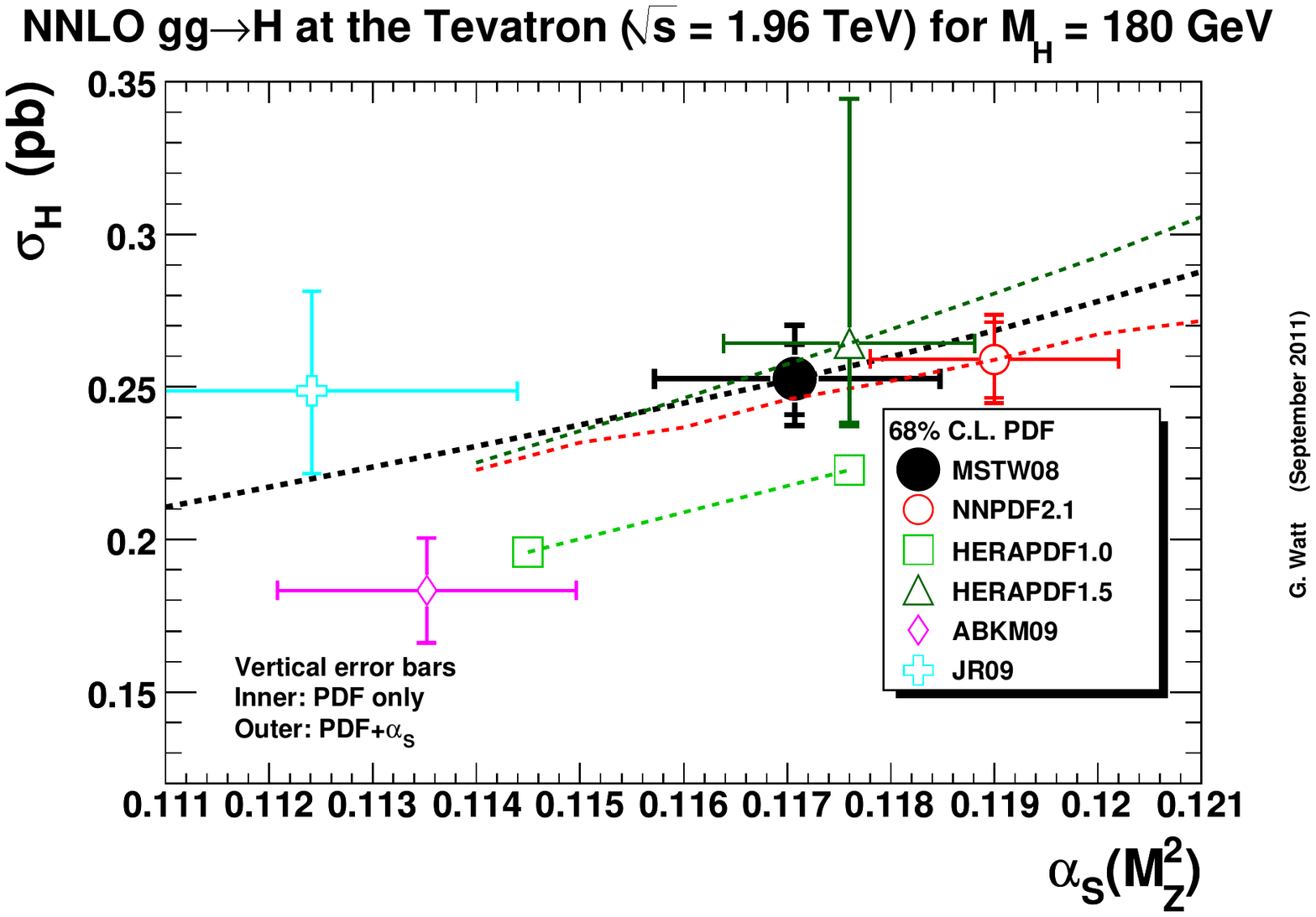}
  \includegraphics[width=0.5\textwidth]{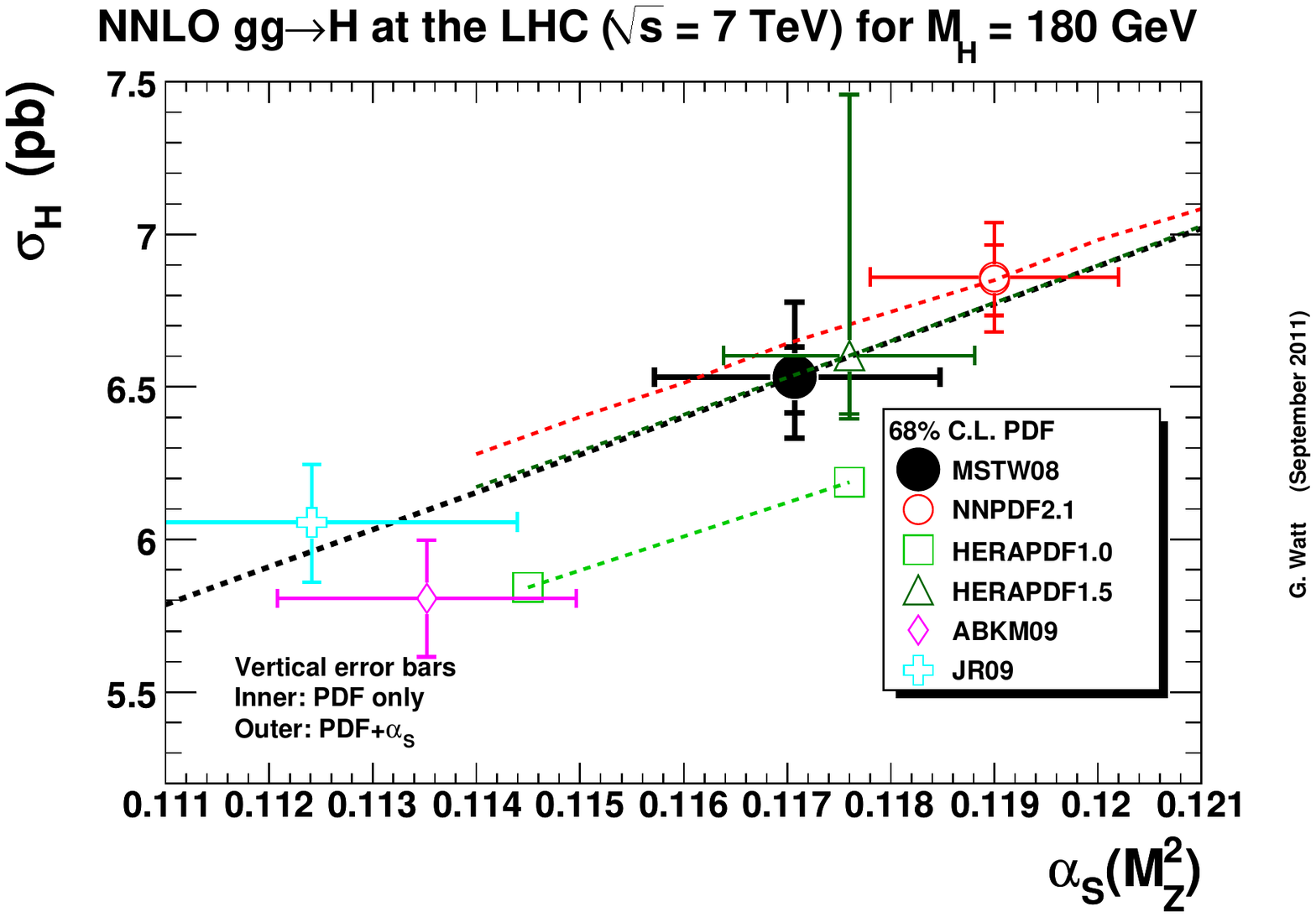}
  \caption{NNLO $gg\to H$ total cross sections, plotted as a function of $\alpha_S(M_Z^2)$, for $M_H=180$~GeV.\label{fig:gghvsasmz}}
\end{figure*}
\begin{figure*}[!t]
  \vspace*{-5cm}
  \includegraphics[width=0.5\textwidth]{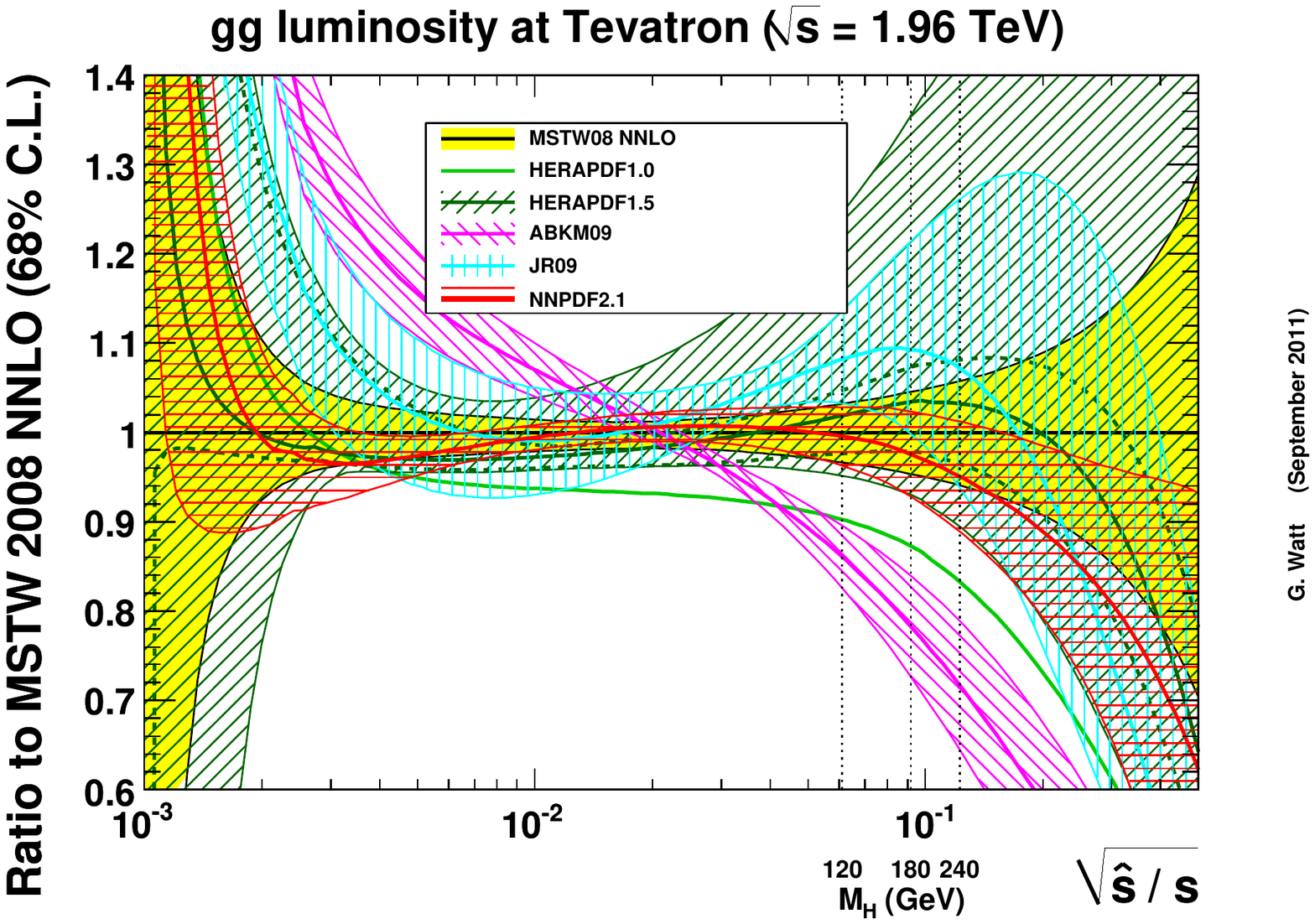}
  \includegraphics[width=0.5\textwidth]{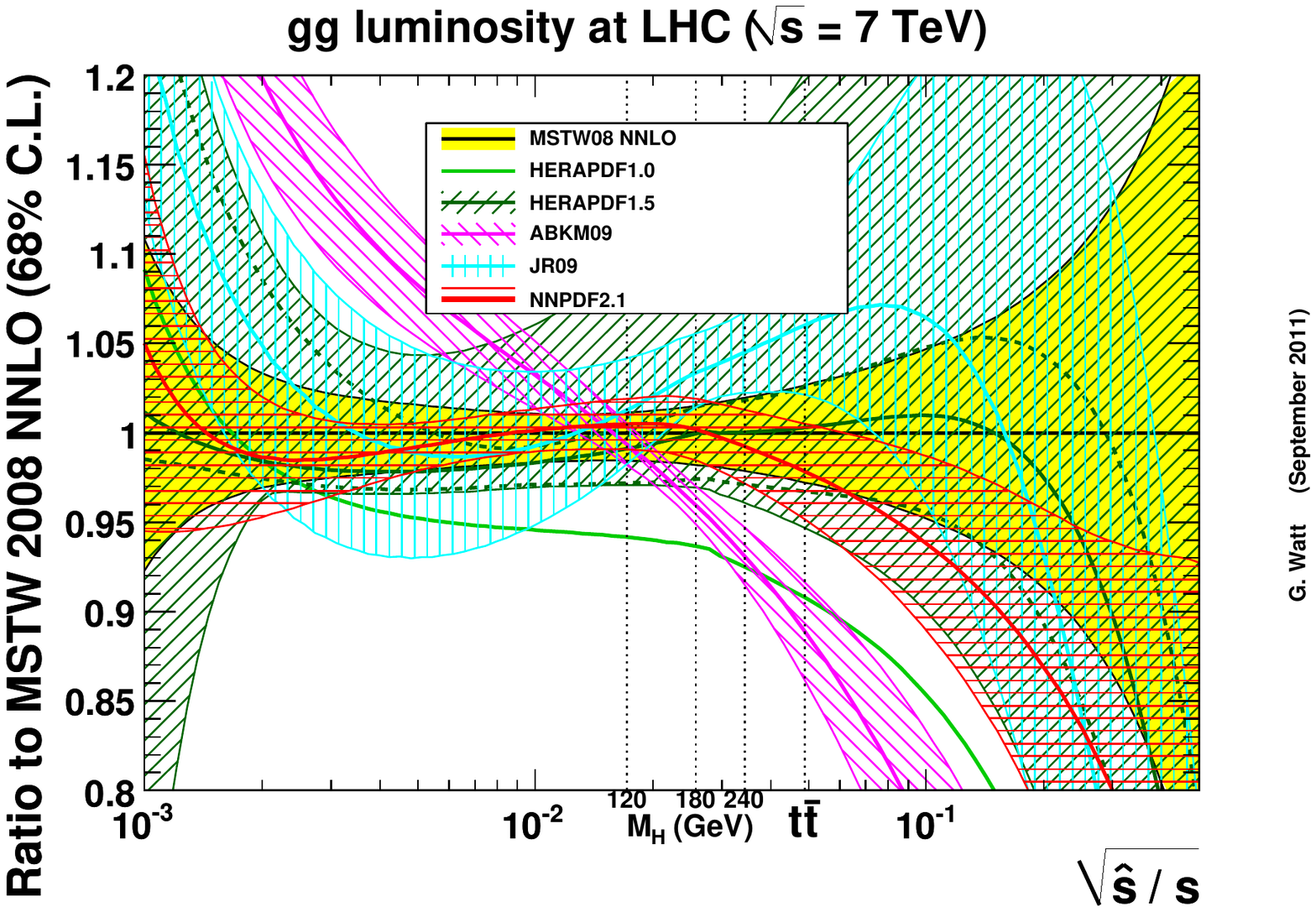}
  \caption{NNLO gluon--gluon luminosities as the ratio with respect to MSTW 2008, plotted as a function of $\sqrt{\hat{s}/s}$.\label{fig:gglumi}}
  \vspace*{1cm}
\end{figure*}

\begin{figure*}[t]
  \vspace*{-5cm}
  \includegraphics[width=0.5\textwidth]{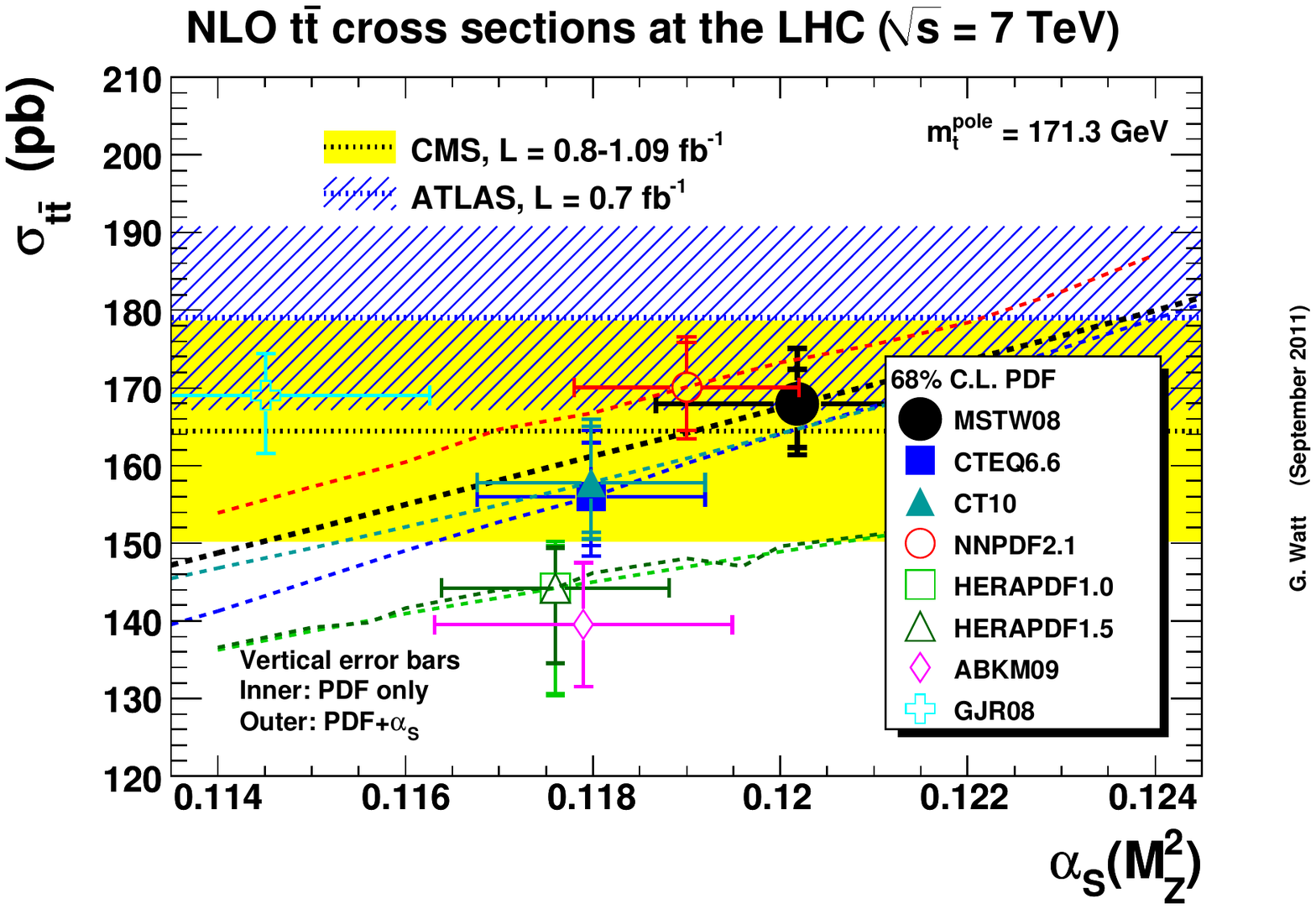}
  \includegraphics[width=0.5\textwidth]{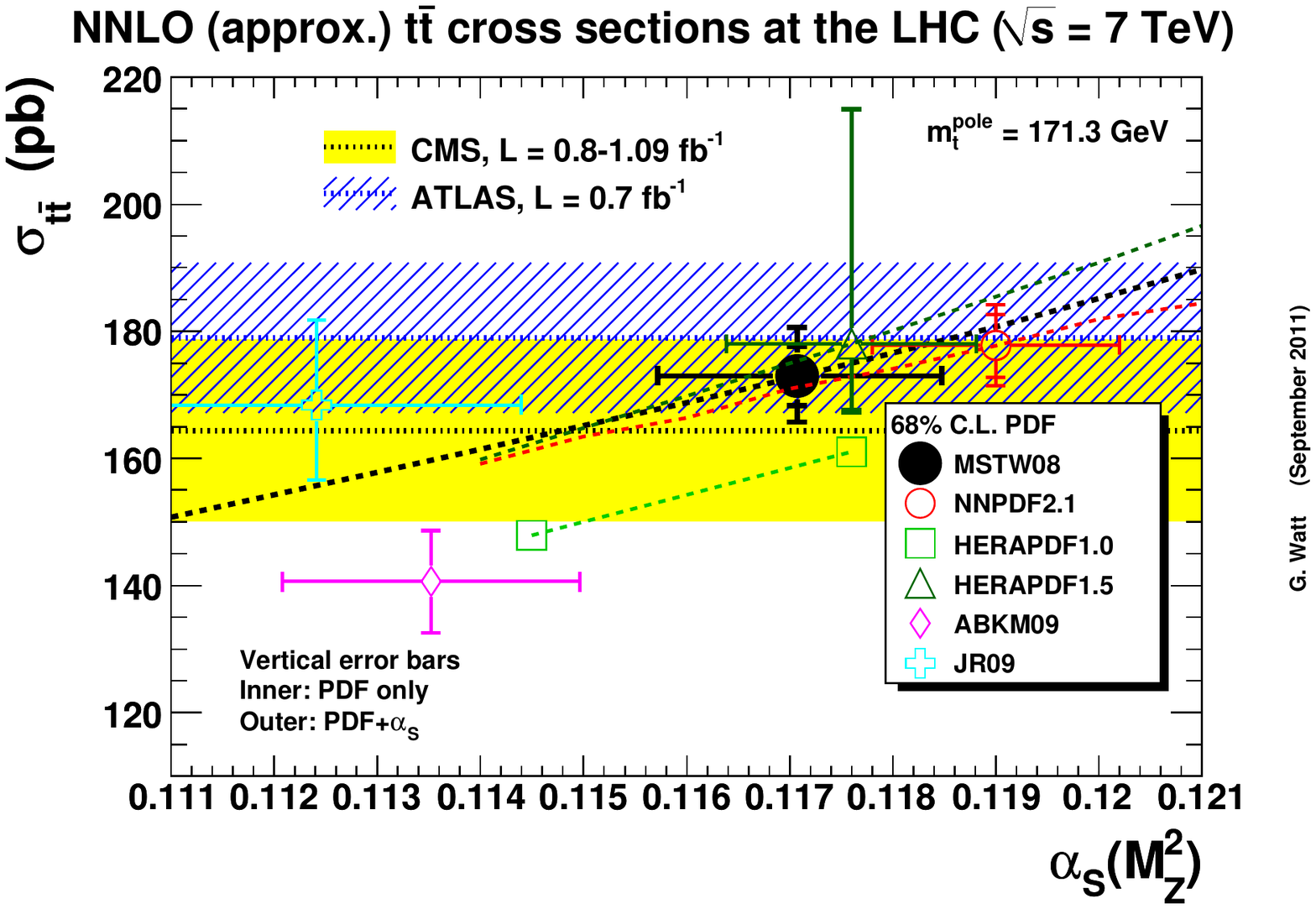}
  \caption{NLO and NNLO (approx.) $t\bar{t}$ total cross sections at the LHC, plotted as a function of $\alpha_S(M_Z^2)$, for $m_t=171.3$~GeV, and compared to the single most precise current LHC measurements from CMS~\cite{CMS:ttbar} and ATLAS~\cite{ATLAS:ttbar}.\label{fig:ttbarvsasmz}}
\end{figure*}

The $gg\to H$ cross sections at the Tevatron and LHC start at $\mathcal{O}(\alpha_S^2)$ at LO, with anomalously large higher-order corrections, therefore they are directly sensitive to the value of $\alpha_S(M_Z^2)$.  Moreover, there is a known correlation between the value of $\alpha_S$ and the gluon distribution, which additionally affects the $gg\to H$ cross sections.  In Fig.~\ref{fig:gghvsasmz} we show this sensitivity by plotting the Higgs cross sections versus $\alpha_S(M_Z^2)$ at the Tevatron and LHC for a Higgs mass $M_H=180$~GeV.  The format of the plots is the same as in Fig.~\ref{fig:wzvsasmz}, but in this case the effect of the additional $\alpha_S$ uncertainty is more sizeable.  It is apparent from the plots that at least part of the MSTW08/ABKM09 discrepancy for Higgs cross sections is due to using quite different values of $\alpha_S(M_Z^2)$ at NNLO, specifically $\alpha_S(M_Z^2) = 0.1135\pm 0.0014$ for ABKM09~\cite{Alekhin:2009ni} compared to $\alpha_S(M_Z^2) = 0.1171\pm 0.0014$ for MSTW08~\cite{Martin:2009iq,Martin:2009bu}.  Comparing cross-section predictions at the same value of $\alpha_S(M_Z^2)$ would reduce the MSTW08/ABKM09 discrepancy at the LHC, but there would still be a significant discrepancy at the Tevatron (see also the later Table~\ref{tab:nmcfl}).

At LO, the PDF dependence of the $gg\to H$ total cross section is simply given by the gluon--gluon luminosity evaluated at a partonic centre-of-mass energy $\sqrt{\hat{s}} = M_H$, i.e.
\[
\frac{\partial {\cal L}_{gg}}{\partial \hat{s}} = \frac{1}{s} \int_\tau^1\frac{{\rm d}x}{x}g(x,\hat{s})g(\tau/x,\hat{s}),
\]
where $g(x,\mu^2=\hat{s})$ is the gluon distribution and $\tau\equiv \hat{s}/s$.  In Fig.~\ref{fig:gglumi} we show the gluon--gluon luminosities calculated using different PDF sets and taken as the ratio with respect to the MSTW 2008 NNLO value, at centre-of-mass energies corresponding to the Tevatron and LHC.  The relevant values of $\sqrt{\hat{s}} = M_{H} = \{120, 180, 240\}$~GeV are indicated, along with the threshold for $t\bar{t}$ production at the LHC, $\sqrt{\hat{s}} = 2m_t$ with $m_t=171.3$~GeV, where this process is predominantly $gg$-initiated at the LHC.  Indeed, $t\bar{t}$ production at the LHC is strongly correlated with $gg\to H$ production at the Tevatron, with both processes probing the gluon distribution at similar $x$ values, as seen from Fig.~\ref{fig:gglumi}.  There is reasonable agreement for the global fits (MSTW08 and NNPDF2.1), but more variation for the other sets, particularly at large $\hat{s}$, where the HERAPDF1.0 set with $\alpha_S(M_Z^2)=0.1176$, and especially the ABKM09 set, has a much softer high-$x$ gluon distribution, and this feature has a direct impact on the $gg\to H$ cross sections, particularly at the Tevatron (see Fig.~\ref{fig:gghvsMH}).  Again, we note that the central value of HERAPDF1.5 is in good agreement with the global fits, but it has a very large uncertainty in the upwards direction, and we will return to this feature later.

More than 80\% of the NLO $t\bar{t}$ cross section comes from the $gg$ channel for the LHC with $\sqrt{s} = 7$~TeV, rising to almost 90\% at $\sqrt{s} = 14$~TeV, compared to less than 15\% at the Tevatron ($\sqrt{s} = 1.96$~TeV).  The significant difference in the initial parton composition for $t\bar{t}$ production is due partly to the lower Tevatron energy ($pp$ collisions at $\sqrt{s}=1.96$~TeV would give around 50\% of the $t\bar{t}$ cross section from the $gg$ channel), but mainly due to the valence--valence nature of the $q\bar{q}\to t\bar{t}$ channel in $p\bar{p}$ collisions.  The partonic subprocess is $\mathcal{O}(\alpha_S^2)$ at LO.  There is therefore a strong dependence on both the gluon distribution (at $x\sim 2m_t/\sqrt{s}=0.05$) and $\alpha_S$.  We calculate $t\bar{t}$ production (without decay) for a top-quark pole mass $m_t=171.3$~GeV (PDG 2009 best value), with a fixed scale choice of $\mu_R=\mu_F=m_t$.  We show the $t\bar{t}$ total cross sections at the LHC ($\sqrt{s} = 7$~TeV), plotted as a function of $\alpha_S(M_Z^2)$, in Fig.~\ref{fig:ttbarvsasmz}, with 68\% C.L.~PDF+$\alpha_S$ uncertainties.  The NNLO calculation of the total cross section for $t\bar{t}$ production is still in progress, although various approximations based on threshold resummation are available.  We use the \textsc{hathor}~\cite{Aliev:2010zk} public code with the default settings for an approximate ``NNLO'' calculation, although we make no attempt to quantify the theoretical uncertainty (other than from PDFs and $\alpha_S$).  A more complete study of the theoretical uncertainties in the approximate NNLO calculation is clearly important, but it is beyond the scope of this work (see Refs.~\cite{Kidonakis:2011ca,Beneke:2011mq,Cacciari:2011hy} for some recent studies).  As a rough indication, the scale uncertainties are estimated to be 13\% at NLO and significantly smaller in the approximate NNLO calculations (for example, 8\% at NLO+NNLL~\cite{Cacciari:2011hy}).  The predicted $t\bar{t}$ cross section has a fairly strong dependence on the assumed top-quark mass $m_t$, such that comparison of the measured cross section with theory predictions even allows an extraction of $m_t$.  As some indication of the $m_t$ dependence, increasing $m_t$ by 2~GeV to give a value close to the current Tevatron average of $m_t=173.2\pm0.9$~GeV~\cite{Lancaster:2011wr} decreases the $t\bar{t}$ cross section at the 7~TeV LHC by about 10~pb (or 6\%) at both NLO and NNLO with MSTW08 PDFs.  Bearing these caveats in mind, we compare to the single most precise current CMS measurement ($e$/$\mu$+jets+$b$-tag) of~\cite{CMS:ttbar}
\begin{figure}[t]
  \vspace*{-5cm}
  \includegraphics[width=0.5\textwidth]{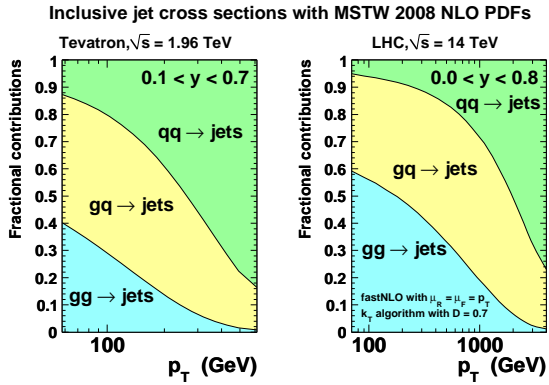}
  \caption{Fractional contributions of the $gg$-, $gq$- and $qq$-initiated processes to inclusive jet production as a function of $p_T$~\cite{Martin:2009bu}.\label{fig:jets_pdffrac}}
\end{figure}
\begin{figure}[t]
  \vspace*{-3cm}
  \includegraphics[width=0.5\textwidth]{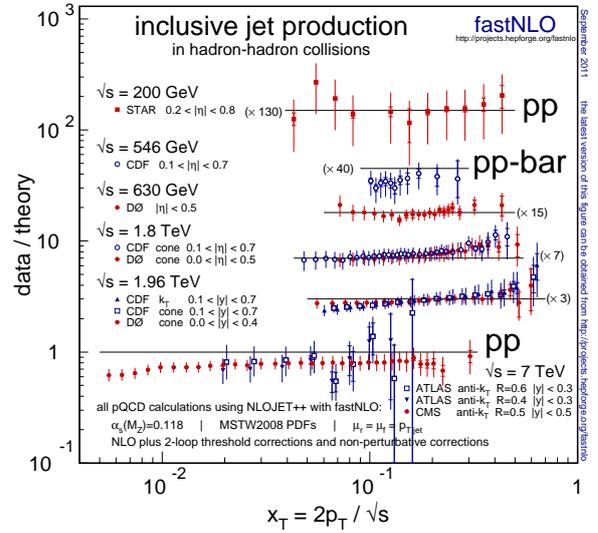}
  \caption{Ratios of data over theory (using MSTW 2008 PDFs) for inclusive jet cross sections at hadron colliders as a function of the scaling variable $x_T=2p_T/\sqrt{s}$.  Plot taken from Ref.~\cite{Wobisch:2011ij}.\label{fig:fastnlo}}
\end{figure}
\[
\hspace*{-5mm}\sigma_{t\bar{t}}=164.4\pm2.8({\rm stat.})\pm11.9({\rm syst.})\pm7.4({\rm lumi.})~{\rm pb},
\]
which is close to a recent CMS combination of $t\bar{t}$ cross-section measurements giving~\cite{CMS:ttbarcomb}
\[
\hspace*{-5mm}\sigma_{t\bar{t}}=165.8\pm2.2({\rm stat.})\pm10.6({\rm syst.})\pm7.8({\rm lumi.})~{\rm pb},
\]
and to the single most precise current ATLAS measurement (using kinematic information of lepton+jets events) of~\cite{ATLAS:ttbar}
\[
\sigma_{t\bar{t}} = 179.0\pm9.8 ({\rm stat.}+{\rm syst.})\pm6.6 ({\rm lumi.})~{\rm pb}.
\]
The approximate NNLO prediction using MSTW08 PDFs shown in Fig.~\ref{fig:ttbarvsasmz} is consistent with both the ATLAS and CMS measurements, while the central value using ABKM09 is almost 2-$\sigma$ below CMS and more than 3-$\sigma$ below ATLAS.  The discrepancy for ABKM09 would increase further if using the more up-to-date value of $m_t=173.2\pm0.9$~GeV~\cite{Lancaster:2011wr} rather than $m_t=171.3$~GeV used in the plots, which would reduce all theory predictions by around 6\%.  The HERAPDF1.0/1.5 sets at NLO are also disfavoured by the LHC data, as is the HERAPDF1.0 NNLO set with $\alpha_S(M_Z^2) = 0.1145$.

It would be a worrying situation if the $t\bar{t}$ cross section at the LHC was needed to discriminate between PDF sets.  The measured $t\bar{t}$ cross section is commonly used to constrain new physics contributions, therefore it is questionable whether it should be used directly to constrain PDFs.  Rather, we would hope that the gluon distribution (and $\alpha_S$) would be sufficiently constrained by other data sets that the $t\bar{t}$ cross section is a \emph{prediction} rather than a direct PDF constraint.  If \emph{only} the ABKM09 PDF set was available, one can only imagine what new physics scenarios could be conjured up to explain the ``excess'' of data over theory seen in Fig.~\ref{fig:ttbarvsasmz}.  

\begin{figure*}[t]
  \fbox{\includegraphics[width=\textwidth]{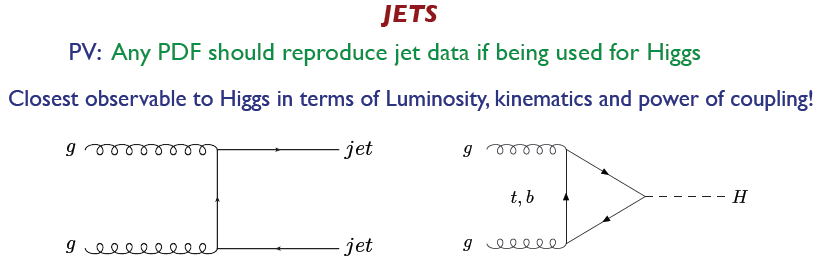}}
  \caption{Jets as a discriminator of the high-$x$ gluon distribution (extract from talk by D.~de Florian~\cite{deFlorian:2011}).\label{fig:deflorian}}
\end{figure*}

Measurements of the scaling violations of DIS structure functions can be used to constrain the small-$x$ gluon, although there is no direct constraint on the large-$x$ gluon from inclusive DIS.  To constrain the high-$x$ gluon distribution we need to look for processes where the gluon appears in the initial state at LO, and the best example is inclusive jet production at hadron colliders.  In Fig.~\ref{fig:jets_pdffrac} we show the composition of the initial state as a function of the jet transverse momentum.  There is a transition from gluon--gluon and gluon--quark at low $p_T$ values to quark--quark at high $p_T$ values, but even at high $p_T$ values the gluon--quark contribution is still significant.  (Here, we do not distinguish between quarks and antiquarks.)  Since the quark distributions can be constrained by other data sets, it means that jet production mainly constrains the gluon distribution.  The plot in Fig.~\ref{fig:fastnlo}, taken from Ref.~\cite{Wobisch:2011ij}, shows data plotted as a function of the scaling variable $x_T=2p_T/\sqrt{s}$, which is related to the $x$ values probed at central rapidity.  The current LHC data are generally at lower $x_T$ than at the Tevatron, so presently the best constraint on the high-$x$ gluon comes from the Tevatron jet data.  Indeed, other than the $t\bar{t}$ cross section, jets are perhaps the closest observable to Higgs in terms of the partonic luminosity, kinematics and power of coupling; see Fig.~\ref{fig:deflorian} taken from Ref.~\cite{deFlorian:2011}.  The crucial question we must address, therefore, is how well do the ``non-global'' fits describe the Tevatron jet data?

We will not consider the less reliable Tevatron Run I data, which prefer a much harder high-$x$ gluon distribution~\cite{Martin:2009iq}, and are obtained using less sophisticated jet algorithms.  The three data sets on inclusive jet production from the Tevatron Run II~\cite{Abulencia:2007ez,Aaltonen:2008eq,Abazov:2008hua} were all found to be compatible~\cite{Martin:2009iq}.  The MSTW 2008 analysis~\cite{Martin:2009iq} included the CDF Run II inclusive jet data using the $k_T$ jet algorithm~\cite{Abulencia:2007ez} and the D{\O} Run II inclusive jet data using a cone jet algorithm~\cite{Abazov:2008hua}.  Consistency was checked with the CDF Run II inclusive jet data using the cone-based Midpoint jet algorithm~\cite{Aaltonen:2008eq}, but this data set was not included in the final MSTW08 fit, since it is essentially the same measurement (using 1.13~fb$^{-1}$) as Ref.~\cite{Abulencia:2007ez} (using 1.0~fb$^{-1}$), differing mainly by the choice of jet algorithm.  The $k_T$ jet algorithm is theoretically preferred due to its property of infrared safety.  We therefore focus here on the CDF Run II inclusive jet data using the $k_T$ jet algorithm~\cite{Abulencia:2007ez}, and the description using NNLO PDFs, but the other Tevatron Run II jet data sets and the description using NLO PDFs are considered in detail in Ref.~\cite{Thorne:2011kq}.

One obvious problem is that the complete NNLO partonic cross section ($\hat{\sigma}$) for inclusive jet production is currently unknown, and needs to be approximated with the NLO $\hat{\sigma}$ supplemented by 2-loop threshold corrections~\cite{Kidonakis:2000gi}.  We calculate jet cross sections using \textsc{fastnlo}~\cite{Kluge:2006xs} (based on \textsc{nlojet++}~\cite{Nagy:2001fj,Nagy:2003tz}), which includes these 2-loop threshold corrections.  Following the usual way of estimating theoretical uncertainties due to unknown higher-order corrections, we take different scale choices $\mu_R=\mu_F=\mu=\{p_T/2,p_T,2p_T\}$ as some indication of the theoretical uncertainty.  Smaller scale choices raise the partonic cross section, so favour softer high-$x$ gluon distributions~\cite{Martin:2009iq}, and the central $\mu=p_T$ was chosen for the final MSTW08 fit~\cite{Martin:2009iq}.  More comments on the scale dependence are given in Ref.~\cite{Thorne:2011kq}.

\begin{table*}[t]
  \begin{center}
    \begin{tabular}{|l|l|l|l|l|}
      \hline
      NNLO PDF & $\alpha_S(M_Z^2)$ & $\mu=p_T/2$ & $\mu=p_T$ & $\mu=2p_T$ \\
      \hline
      MSTW08 & 0.1171 & 1.39 ($+$0.35) & 0.69 ($-$0.45) & 0.97 ({\it$-$1.30}) \\
      NNPDF2.1 & 0.1190 & 0.68 ($-$0.77) & 0.71 ({\it$-$2.02}) & 0.71 ({\bf$-$3.46}) \\
      HERAPDF1.0 & 0.1145 & 2.37 ({\it$-$2.65}) & 1.48 ({\bf$-$3.64}) & 1.29 ({\bf$-$4.12}) \\
      HERAPDF1.0 & 0.1176 & 2.24 ($-$0.48) & 1.13 ({\it$-$1.60}) & 1.09 ({\it$-$2.23}) \\
      HERAPDF1.5 & 0.1176 & 1.61 ({\it$+$1.22}) & 0.77 ($+$0.30) & 1.06 ($-$0.39) \\
      ABKM09 & 0.1135 & 1.53 ({\bf$-$4.27}) & 1.23 ({\bf$-$5.05}) & 1.44 ({\bf$-$5.65}) \\
      JR09 & 0.1124 & 0.75 ($+$0.13) & 1.26 ($-$0.61) & 2.20 ({\it$-$1.22}) \\
      \hline
    \end{tabular}
  \end{center}
  \caption{Values of $\chi^2/N_{\rm pts.}$ for the CDF Run II inclusive jet data using the $k_T$ jet algorithm~\cite{Abulencia:2007ez} with $N_{\rm pts.}=76$ and $N_{\rm corr.}=17$, for different NNLO PDF sets and different scale choices $\mu_R=\mu_F=\mu=\{p_T/2,p_T,2p_T\}$.  No restriction is imposed on the shift in normalisation and the optimal value of ``$-r_{\rm lumi.}$'' is shown in brackets, where the data points are shifted as $D_i\to D_i(1-0.058\,r_{\rm lumi.})$.  Values of $|r_{\rm lumi.}|\in[1,3]$ are shown in \emph{italics} and values $|r_{\rm lumi.}|>3$ are shown in \textbf{bold}.\label{tab:cdfkt_freenorm}}
\end{table*}

It is important to account for \emph{correlated} systematic uncertainties of the experimental data points.  The full correlated error information is accounted for by using a goodness-of-fit ($\chi^2$) definition given by~\cite{Stump:2001gu,Pumplin:2002vw}
\begin{equation} \label{eq:chisqcorr}
  \chi^2 \;=\; \sum_{i=1}^{N_{\rm pts.}} \left(\frac{\hat{D}_{i}-T_{i}}{\sigma_{i}^{\rm uncorr.}}\right)^2 \;+\; \sum_{k=1}^{N_{\rm corr.}}r_{k}^2,
\end{equation}
where $T_{i}$ are the theory predictions and $\hat{D}_{i} \equiv D_{i} - \sum_{k=1}^{N_{\rm corr.}}r_{k}\,\sigma_{k,i}^{\rm corr.}$ are the data points allowed to shift by the systematic errors in order to give the best fit.  Here, $i=1,\ldots,N_{\rm pts.}$ labels the individual data points and $k=1,\ldots,N_{\rm corr.}$ labels the individual correlated systematic errors.  The data points $D_{i}$ have uncorrelated (statistical and systematic) errors $\sigma_{i}^{\rm uncorr.}$ and correlated systematic errors $\sigma_{k,i}^{\rm corr.}$.  The optimal shifts of the data points by the systematic errors, $r_{k}$, are solved for analytically by minimising the $\chi^2$ in Eq.~(\ref{eq:chisqcorr}).  There is a clear trade-off between the systematic shifts $r_k$ and the parameters of the gluon distribution.  Deficiencies in the theory calculation can be masked to some extent by large systematic shifts, therefore it is important to check that the optimal $r_k$ values are not unreasonable.  This is straightforward when using a $\chi^2$ definition like Eq.~(\ref{eq:chisqcorr}), but is more difficult using an equivalent form
\begin{equation} \label{eq:chisqcov}
  \chi^2 \;=\; \sum_{i=1}^{N_{\rm pts.}}\sum_{i^\prime=1}^{N_{\rm pts.}}(D_i-T_i)\left(V^{-1}\right)_{ii^\prime}(D_{i^\prime}-T_{i^\prime}),
\end{equation}
written in terms of the inverse of the experimental covariance matrix,
\[
  V_{ii^\prime} \;=\; \delta_{ii^\prime}\,(\sigma_{i}^{\rm uncorr.})^2\;+\;\sum_{k=1}^{N_{\rm corr.}}\sigma_{k,i}^{\rm corr.}\,\sigma_{k,i^\prime}^{\rm corr.},
\]
as used by the ABKM and NNPDF fitting groups.  More precisely, NNPDF use a refinement to treat normalisation errors as multiplicative~\cite{Ball:2009qv}, while Alekhin (ABKM) treats all correlated systematic errors as multiplicative~\cite{Alekhin:1996za,Alekhin:2000ch}.

In Table~\ref{tab:cdfkt_freenorm} we give the $\chi^2$ per data point, calculated using Eq.~(\ref{eq:chisqcorr}), for the CDF Run II data on inclusive jet production using the $k_T$ jet algorithm~\cite{Abulencia:2007ez}, for different NNLO PDF sets and different scale choices $\mu_R=\mu_F=\mu=\{p_T/2,p_T,2p_T\}$, where $p_T$ is the jet transverse momentum.  We treat the luminosity uncertainty as any other correlated systematic.  However, we find that the relevant systematic shift $r_{\rm lumi.}\sim 3$--$5$ for some PDF sets with soft high-$x$ gluon distributions (e.g.~ABKM09 and HERAPDF1.0), which is clearly completely unreasonable, as it means that the data points are normalised downwards by 3--5 times the nominal luminosity uncertainty (5.8\% for CDF).  The penalty term $r_{\rm lumi.}^2$ will contribute only 9--25 units to the total $\chi^2$ given by Eq.~(\ref{eq:chisqcorr}), which can therefore still lead to reasonably low overall $\chi^2$ values.

It is the usual situation at collider experiments that the luminosity determination is common to all cross sections measured from a given data set, so the requirement of a single common luminosity is mandatory when fitting multiple measurements taken during a single running period.  All NNLO PDFs are in good agreement with the Tevatron $W$ and $Z$ cross sections.  If the Tevatron jet data were normalised downwards by 20--30\% (i.e.~3--5 times the luminosity uncertainty), the Tevatron $W$ and $Z$ total cross sections would need to normalised downwards by the same amount, resulting in complete disagreement with all theory predictions.  This example illustrates the utility of simultaneously fitting $W$ and $Z$ cross sections together with jet cross sections at the Tevatron (and LHC).  The luminosity shifts, common to both data sets, are effectively determined by the more precise $W$ and $Z$ cross sections.  The luminosity uncertainty is then effectively removed from the jet cross sections, thereby allowing the jet data to provide a tighter constraint on the gluon distribution (and $\alpha_S$).

\begin{table*}[t]
    \begin{center}
      \begin{tabular}{|l|l|l|l|l|}
        \hline
        NNLO PDF & $\alpha_S(M_Z^2)$ & $\mu=p_T/2$ & $\mu=p_T$ & $\mu=2p_T$ \\
        \hline
        MSTW08 & 0.1171 & 1.39 ($+$0.35) & {\bf 0.69} ($-$0.45) & 0.97 ($-$1.00) \\
        NNPDF2.1 & 0.1190 & {\bf 0.68} ($-$0.77) & {\bf 0.81} ($-$1.00) & 1.29 ($-$1.00) \\
        HERAPDF1.0 & 0.1145 & 2.64 ($-$1.00) & 2.15 ($-$1.00) & 2.20 ($-$1.00) \\
        HERAPDF1.0 & 0.1176 & 2.24 ($-$0.48) & 1.17 ($-$1.00) & 1.23 ($-$1.00) \\
        HERAPDF1.5 & 0.1176 & 1.61 ($+$1.00) & {\bf 0.77} ($+$0.30) & 1.06 ($-$0.39) \\
        ABKM09 & 0.1135 & 2.55 ($-$1.00) & 2.76 ($-$1.00) & 3.41 ($-$1.00) \\
        JR09 & 0.1124 & {\bf 0.75} ($+$0.13) & 1.26 ($-$0.61) & 2.21 ($-$1.00) \\
        \hline
      \end{tabular}
    \end{center}
    \caption{Same as Table~\ref{tab:cdfkt_freenorm}, but at most a 1-$\sigma$ shift in normalisation is allowed.  The optimal value of ``$-r_{\rm lumi.}$'' is shown in brackets, subject to this restriction, where the data points are shifted as $D_i\to D_i(1-0.058\,r_{\rm lumi.})$.  We highlight in bold those values lying inside the 90\% C.L.~region, defined by Eq.~(\ref{eq:90percentCL}), which gives $\chi^2/N_{\rm pts.} < 0.83$.\label{tab:cdfkt}}
\end{table*}

To avoid these completely unrealistic luminosity shifts, $r_{\rm lumi.}\sim 3$--$5$, without going into the complication of simultaneously including $W$ and $Z$ cross sections in the $\chi^2$ computation, we will calculate the $\chi^2$ values for the Tevatron jet data using Eq.~(\ref{eq:chisqcorr}), but with the simple restriction that the relevant systematic shift $|r_{\rm lumi.}|\le 1$.  More practically, this means that if $|r_{\rm lumi.}|>1$ for any particular PDF set, we fix $r_{\rm lumi.}$ at $\pm1$ and reevaluate Eq.~(\ref{eq:chisqcorr}) with the luminosity removed from the list of correlated systematics.  The results are given in Table~\ref{tab:cdfkt}.  We highlight in bold the $\chi^2$ values lying inside the 90\% C.L.~region defined as
\begin{equation} \label{eq:90percentCL}
  \chi^2 < \left(\frac{\chi_{0}^2}{\xi_{50}}\right)\xi_{90},
\end{equation}
where $\xi_{50}$ and $\xi_{90}$ are the 50th and 90th percentiles of the $\chi^2$-distribution with $N_{\rm pts.}=76$ degrees of freedom.  (These quantities are defined in detail in Sect.~6.2 of Ref.~\cite{Martin:2009iq}.)  Here, $\chi_{0}^2$ is defined as the lowest $\chi^2$ value of all theory predictions, i.e.~assumed to be close to the best possible fit, so that the rescaling factor $\chi_{0}^2/\xi_{50}$ in Eq.~(\ref{eq:90percentCL}) empirically accounts for any unusual fluctuations preventing the best possible fit having $\chi^2 \simeq \xi_{50} \simeq N_{\rm pts.}$~\cite{Stump:2001gu}.  The 90\% C.L.~region given in this way is used to determine the PDF uncertainties according to the ``dynamical tolerance'' prescription introduced in Ref.~\cite{Martin:2009iq}, so PDF sets with $\chi^2$ values far outside this region cannot be considered to give an acceptable description of the data.  We see from Table~\ref{tab:cdfkt} that MSTW08, NNPDF2.1 and HERAPDF1.5 give an acceptable description for $\mu=p_T$, while HERAPDF1.0 (with the lower $\alpha_S$ value) and ABKM09 give $\chi^2/N_{\rm pts.}\sim 2$--$3$.  The JR09 set and the HERAPDF1.0 set with $\alpha_S(M_Z^2) = 0.1176$ give a better description, and give predictions for $gg\to H$ cross sections at the Tevatron which are closer to the MSTW08 predictions than those from ABKM09 and the HERAPDF1.0 NNLO set with $\alpha_S(M_Z^2) = 0.1145$.  The same trend is apparent, but to a somewhat lesser extent, for the CDF Run II inclusive jet data using the cone-based Midpoint jet algorithm~\cite{Aaltonen:2008eq} and the D{\O} Run II inclusive jet data using a cone jet algorithm~\cite{Abazov:2008hua}; see Ref.~\cite{Thorne:2011kq} for the details and more discussion.

To summarise, comparison with Tevatron jet data is subtle because of the large correlated systematic uncertainties.  The systematic shifts can compensate for inadequacies in the theory calculation.  The traditional $\chi^2$ definition in terms of the experimental covariance matrix, Eq.~(\ref{eq:chisqcov}), can hide such systematic shifts.  In particular, we find that the Tevatron jet data need to be normalised downwards by typically between 3-$\sigma$ and 5-$\sigma$ (see Table~\ref{tab:cdfkt_freenorm}) to achieve the best agreement with some PDF sets, particularly the ABKM09 predictions.  Even if the luminosity shift is artificially constrained, the other systematic shifts move by large amounts for the inclusive jet data, incompatible with the Gaussian expectation.  No such problems are observed for the MSTW08 predictions.  It can also be seen from the plots in Ref.~\cite{Alekhin:2011cf} that the unshifted Tevatron jet data lie significantly above the theory predictions even after including these data in variants of the ABKM09 fit.  Constraining the Tevatron luminosity shifts, for example, so that the predicted $W$ and $Z$ cross sections agreed with Tevatron data, would increase the constraining power of the Tevatron jet data and thereby very likely give a larger $\alpha_S$ and high-$x$ gluon distribution than the current ABM studies~\cite{Alekhin:2011cf}.  Even with the existing treatment, the NNLO Tevatron $gg\to H$ cross section for $M_H=165$~GeV goes up by 15\% when including the CDF Run II ($k_T$ jet algorithm)~\cite{Abulencia:2007ez} data set in a variant of the ABKM09 fit~\cite{Alekhin:2011cf}.

\section{Strong coupling $\alpha_S$ from DIS}
\label{sec:alphaS}

A recent claim has been made~\cite{Alekhin:2011ey} that the bulk of the MSTW08/ABKM09 difference in both the extracted $\alpha_S(M_Z^2)$ value and the $gg\to H$ predictions is explained by the treatment of NMC data~\cite{Arneodo:1996qe}.  The differential cross section for DIS of charged leptons off nucleons, $\ell N\to \ell X$, neglecting the nucleon and lepton masses, and assuming single-photon exchange, is
\begin{equation}
  \label{eq:d2sigma}
  \hspace*{-5mm}\frac{{\rm d}^2\sigma}{{\rm d}x\,{\rm d}Q^2} \simeq \frac{4\pi\alpha^2}{x\,Q^4}\left[1-y+\frac{y^2/2}{1+R(x,Q^2)}\right]F_2(x,Q^2),
\end{equation}
where $R=\sigma_L/\sigma_T\simeq F_L/(F_2-F_L)$ is the ratio of the $\gamma^* N$ cross sections for longitudinally and transversely polarised photons, $Q^2$ is the photon virtuality, $x$ is the Bjorken variable and $y\simeq Q^2/(x\,s)$ is the inelasticity (with $\sqrt{s}$ the $\ell N$ centre-of-mass energy).  The ABKM09~\cite{Alekhin:2009ni} analysis fitted the NMC differential cross sections directly, calculating $F_L$ to $\mathcal{O}(\alpha_S^2)$ and including empirical higher-twist corrections.  The MSTW08~\cite{Martin:2009iq} analysis instead fitted the NMC $F_2$ values corrected for $R$, where $R(x,Q^2) = R_{\rm NMC}(x)$ if $x<0.12$ or $R(x,Q^2) = R_{1990}(x,Q^2)$ if $x>0.12$~\cite{Arneodo:1996qe}.  Here, $R_{\rm NMC}(x)$ was a ($Q^2$-independent) value extracted from NMC data, while $R_{1990}(x,Q^2)$ was a $Q^2$-dependent empirical parameterisation of SLAC data dating from 1990~\cite{Whitlow:1990gk}.  By replacing the NMC differential cross-section data by NMC $F_2$ data, ABM~\cite{Alekhin:2011ey} found that their best-fit $\alpha_S(M_Z^2)$ moved from 0.1135 to 0.1170 and their $gg\to H$ cross sections at the Tevatron and LHC moved closer to the MSTW08 values.  ABM~\cite{Alekhin:2011ey} therefore concluded that the use of NMC $F_2$ data in the MSTW08 fit rather than the differential cross section is the main reason for the higher $\alpha_S(M_Z^2)$ and Higgs cross sections obtained with MSTW08.

\begin{figure}[t]
  \vspace*{-5cm}
  \includegraphics[width=0.5\textwidth]{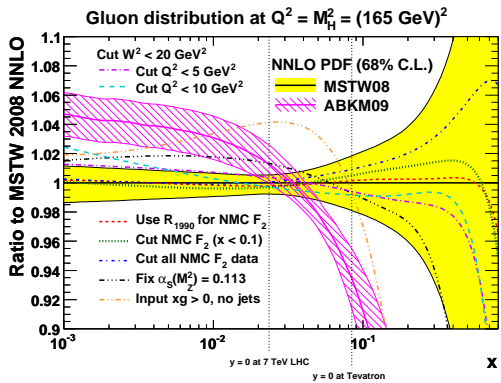}
  \caption{Effect of NMC treatment on the gluon distribution at a scale $Q^2=(165~{\rm GeV})^2$.  The values of $x=M_H/\sqrt{s}$ relevant for central production (assuming $p_T^H=0$) of a SM Higgs boson of mass $M_H=165$~GeV at the Tevatron and LHC are indicated.  We also show the effect of raising the cuts imposed on the DIS data compared to the default of removing data with $Q^2<2$~GeV$^2$ and $W^2<15$~GeV$^2$.  Finally, we show the effect of simply fixing $\alpha_S(M_Z^2)$ to be close to the ABKM09 value, or performing a fit with a positive-definite input gluon distribution and no jet data, and we compare directly to ABKM09.  Plot taken from Ref.~\cite{Thorne:2011kq}.\label{fig:nmcfl}}
\end{figure}

\begin{table}[t]
  \centering
  \begin{tabular}{|l|l|}
    \hline
    ABKM09 & MSTW08 \\ \hline
    Fit NMC cross section & Fit NMC $F_2$ \\
    Fit $Q^2\ge 2.5~{\rm GeV}^2$ & Fit $Q^2\ge 2~{\rm GeV}^2$ \\
    Fit $W^2\ge 3.24~{\rm GeV}^2$ & Fit $W^2\ge 15~{\rm GeV}^2$ \\
    Fit higher-twist & Neglect higher-twist \\
    Separated energies $E_\mu$ & Averaged energies $E_\mu$ \\
    Correlated systematics & Neglect correlations \\
    3 gluon parameters & 7 gluon parameters \\
    \textbf{No jet data} & \textbf{Tevatron jet data} \\
    \hline
  \end{tabular}
  \caption{Main differences in the treatment of NMC data by the ABKM09~\cite{Alekhin:2009ni} and MSTW08~\cite{Martin:2009iq} fits.\label{tab:abkm09diff}}
\end{table}
It is certainly more consistent to fit directly to the NMC differential cross-section data, and so this rather dramatic assertion made by ABM~\cite{Alekhin:2011ey} would obviously be very worrying if correct.  However, there are many other differences between the two analyses besides the treatment of $F_L$ for the NMC data, with some relevant differences given in Table~\ref{tab:abkm09diff} and where the last row (inclusion of jet data) is highlighted as being the most important.  Nevertheless, we carried out a detailed investigation of the sensitivity to NMC data in Ref.~\cite{Thorne:2011kq}.  Rather than repeat the MSTW08 analysis by fitting the NMC differential cross sections, we noted that the original NMC paper~\cite{Arneodo:1996qe} made an alternative extraction of $F_2$ values using the SLAC $R_{1990}$ parameterisation~\cite{Whitlow:1990gk}.  We observed that the MSTW08 NNLO prediction, with $F_L$ calculated to $\mathcal{O}(\alpha_S^3)$ and without any higher-twist corrections, gives a good description of the SLAC $R_{1990}$ parameterisation, demonstrating that fitting the alternative NMC $F_2$ data extracted using the SLAC $R_{1990}$ parameterisation will give very similar results to fitting the NMC differential cross sections.
\begin{table*}[t]
  \centering
  \begin{tabular}{|l|c|c|c|}
    \hline
    NNLO PDF & $\alpha_S(M_Z^2)$ & $\sigma_H$ at Tevatron & $\sigma_H$ at 7 TeV LHC\\ \hline
    {\bf MSTW08} & {\bf 0.1171} & {\bf 0.342~pb} & {\bf 7.91~pb} \\ \hline
    Use $R_{1990}$ for NMC $F_2$ & $0.1167$ & $-0.7\%$ & $-0.9\%$ \\
    Cut NMC $F_2$ ($x<0.1$) & $0.1162$ & $-1.2\%$ & $-2.1\%$ \\
    Cut all NMC $F_2$ data & $0.1158$ & $-0.7\%$ & $-2.1\%$ \\ \hline
    Cut $Q^2<5$~GeV$^2$, $W^2<20$~GeV$^2$ & $0.1171$ & $-1.2\%$ & $+0.4\%$ \\
    Cut $Q^2<10$~GeV$^2$, $W^2<20$~GeV$^2$ & $0.1164$ & $-3.0\%$ & $-1.7\%$ \\ \hline
    Fix $\alpha_S(M_Z^2)$ & $0.1130$ & $-11\%$ & $-7.6\%$ \\ 
    Input $xg>0$, no jets & $0.1139$ & $-17\%$ & $-4.9\%$ \\ \hline
    ABKM09 & 0.1135 & $-26\%$ & $-11\%$ \\
    \hline
  \end{tabular}
  \caption{Effect of NMC treatment on $\alpha_S(M_Z^2)$ and Higgs cross sections ($M_H=165$~GeV).  We also show the effect of raising the cuts imposed on the DIS data compared to the default of removing data with $Q^2<2$~GeV$^2$ and $W^2<15$~GeV$^2$.  Finally, we show the effect of simply fixing $\alpha_S(M_Z^2)$ to be close to the ABKM09 value, or performing a fit with a positive-definite input gluon distribution and no jet data, and we compare directly to ABKM09.  Table taken from Ref.~\cite{Thorne:2011kq}.\label{tab:nmcfl}}
\end{table*}
In Table~\ref{tab:nmcfl} we show the effect of repeating the MSTW08 NNLO fit with the NMC $F_2$ data extracted using $R_{1990}$ on $\alpha_S(M_Z^2)$ and the Higgs cross sections (for $M_H=165$~GeV) at the Tevatron and LHC, and in Fig.~\ref{fig:nmcfl} we show the change in the gluon distribution at the corresponding scale.  We make other fits either cutting the NMC $F_2$ data for $x<0.1$, above which the $R$ correction in Eq.~(\ref{eq:d2sigma}) is very small indeed, or completely removing all NMC $F_2$ data.  In all cases there is very little change in $\alpha_S(M_Z^2)$, the gluon distribution, and the Higgs cross section.  We conclude that the treatment of NMC data cannot explain the difference between the MSTW08 and ABKM09 results.  Similar stability has been found by the NNPDF group~\cite{NNPDF:2011we}, but in a somewhat less relevant study with fixed $\alpha_S$.

The cuts on DIS data are not explicitly given in the ABKM09 paper~\cite{Alekhin:2009ni}, but the previous AMP06 paper~\cite{Alekhin:2006zm} mentions that DIS data are removed with $Q^2<2.5$~GeV$^2$ and $W^2<(1.8~{\rm GeV})^2=3.24~{\rm GeV}^2$, compared to the MSTW08 fit which removes DIS data with $Q^2<2$~GeV$^2$ and $W^2<15~{\rm GeV}^2$.  The much weaker cut on the hadronic invariant mass (squared), $W^2\simeq Q^2(1/x-1)$, clearly explains why higher-twist corrections are more important in the ABKM09 analysis.  To investigate the possible effect of neglected higher-twist corrections on the MSTW08 NNLO fit we raised the cuts to remove DIS data with $W^2<20$~GeV$^2$ and either $Q^2<5$~GeV$^2$ or $Q^2<10$~GeV$^2$.  The results are shown in Table~\ref{tab:nmcfl} and Fig.~\ref{fig:nmcfl}.  The changes in $\alpha_S$, the gluon distribution and the Higgs cross sections are generally small and within uncertainties, although with the strongest $Q^2$ cut there is no data constraint below $x=10^{-4}$ and little just above, so the PDFs differ but have large uncertainties at low $x$ values.

In Table~\ref{tab:nmcfl} and Fig.~\ref{fig:nmcfl} we show the results of the MSTW08 NNLO fit with a fixed $\alpha_S(M_Z^2) = 0.113$~\cite{Martin:2009bu} (slightly below the ABKM09 value), and even in this case the gluon distribution and Higgs cross sections move only part of the way towards the ABKM09 result, as already seen in Fig.~\ref{fig:gghvsasmz}.  The MSTW08 NNLO input gluon parameterisation~\cite{Martin:2009iq} has 7 free parameters compared to only 3 for ABKM09, only 2 for JR09 and HERAPDF1.0 (although the value of $Q_0^2$ is optimised in the case of JR09), and only 4 for HERAPDF1.5 NNLO.  In the lack of any direct data constraint on the high-$x$ gluon distribution, the other fits are therefore constrained by the form of the input parameterisation, avoiding potential pathological behaviour such as a negative high-$x$ gluon distribution.  In an attempt to mimic the ABKM09 fit we performed a variant of the MSTW08 NNLO fit without jet data and with the input gluon distribution forced to be positive.  The results of this fit are shown in Table~\ref{tab:nmcfl} and Fig.~\ref{fig:nmcfl} and it goes some way towards reproducing the high-$x$ gluon of the ABKM09 fit and the corresponding Tevatron $gg\to H$ prediction, certainly closer than we come with other modifications.

Other differences between the two analyses are that ABKM09 used the NMC data for separate muon beam energies, whereas MSTW08 used the NMC data averaged over beam energies, which reduces the maximum effect of the change in $R$ for a particular data point, i.e.~at a given $x$ and $Q^2$, a data point at high $y$, and so very sensitive to $R$ at a low beam energy, is at lower $y$ for a higher beam energy.  In the case of the averaged NMC data, correlated systematic uncertainties are unavailable, so the MSTW08 fit simply added errors (other than normalisation) in quadrature.  As with the Tevatron jet data, deficiencies in the theory calculation may be hidden, without much trace, by large systematic shifts implicit in the $\chi^2$ definition, Eq.~(\ref{eq:chisqcov}), similar to that used in the ABKM09 analysis.  We conclude that the greater sensitivity to the treatment of NMC data found by ABM~\cite{Alekhin:2011ey} is due to a variety of reasons, but perhaps most significantly, the inclusion of higher-twist corrections due to the weaker cuts on DIS data, and, as we have repeatedly emphasised, the lack of additional constraints provided by the Tevatron jet data to pin down the high-$x$ gluon distribution.

Note from Figs.~\ref{fig:gghvsMH}, \ref{fig:gghvsasmz}, \ref{fig:gglumi} and \ref{fig:ttbarvsasmz} that the HERAPDF1.5 NNLO prediction has a very large uncertainty in the upwards direction.  For example, the uncertainties on the $t\bar{t}$ cross section can be broken down as
\[
\hspace*{-5mm}\sigma_{t\bar{t}} = 178^{+4}_{-6}({\rm exp.})^{+37}_{-6}({\rm model})^{+0}_{-6}({\rm param.})\pm6(\alpha_S)~{\rm pb}.
\]
The dominant model uncertainty comes from varying the minimum $Q^2$ cut from the default value of $Q^2_{\rm min} = 3.5~{\rm GeV}^2$.  Lowering the cut to $Q^2_{\rm min} = 2.5~{\rm GeV}^2$ increases $\sigma_{t\bar{t}}$ by 9~pb, while raising the cut to $Q^2_{\rm min} = 5~{\rm GeV}^2$ increases $\sigma_{t\bar{t}}$ by 35~pb, almost all of the total uncertainty on $\sigma_{t\bar{t}}$ in the upwards direction.  It is maybe worrying that the NLO version does not exhibit the same sensitivity, perhaps because of the more restrictive parameterisation at NLO (only 2 gluon parameters) than at NNLO (4 gluon parameters).  As shown in Fig.~\ref{fig:nmcfl}, the high-$x$ gluon in the MSTW08 fit is relatively insensitive to raising $Q^2_{\rm min} = 2\to\{5,10\}~{\rm GeV}^2$, primarily because the Tevatron jet data stabilise the fit and so lessen sensitivity to the fine details of the treatment of the DIS data.

There is a common lore (see, for example, Ref.~\cite{Alekhin:2011gj}) that DIS-only fits prefer low $\alpha_S(M_Z^2)$ values, but Ref.~\cite{Martin:2009bu} showed that not all DIS data sets prefer low $\alpha_S(M_Z^2)$ values.  In particular, this was found to be true only for BCDMS data, and for E665 and SLAC $ep$ data, while NMC, SLAC $ed$ and HERA data preferred high $\alpha_S(M_Z^2)$ values within the context of the global fit~\cite{Martin:2009bu}.  See also the recent NNPDF studies using an ``unbiased'' PDF parameterisation~\cite{Lionetti:2011pw,Ball:2011us}.  It is well known that $\alpha_S$ is highly \emph{anticorrelated} with the low-$x$ gluon distribution through scaling violations of HERA data: $\partial F_2/\partial\ln(Q^2)\sim \alpha_S\,g$.  Then $\alpha_S$ is \emph{correlated} with the high-$x$ gluon distribution through the momentum sum rule; see, for example, Fig.~14(b) of Ref.~\cite{Martin:2009bu}.  Restrictive gluon parameterisations, without the negative small-$x$ term allowed by MSTW~\cite{Martin:2009iq}, can therefore bias the extracted $\alpha_S$ value.  For example, the default MSTW08 NNLO fit obtained $\alpha_S(M_Z^2) = 0.1171\pm 0.0014$, while imposing the restriction of a positive input gluon at $Q_0^2=1$~GeV$^2$ gave a best-fit $\alpha_S(M_Z^2) = 0.1157$, but with a $\chi^2$ worse by 63 units for the global fit to 2615 data points.

What is $\alpha_S$ from only DIS data in the MSTW08 NNLO fit?  Recall that the global fit gave $\alpha_S(M_Z^2) = 0.1171\pm 0.0014$~\cite{Martin:2009bu}.  To expand on the studies made in Ref.~\cite{Martin:2009bu}, we performed a new NNLO DIS-only fit, which gave a best-fit $\alpha_S(M_Z^2) = 0.1104$, but with an input gluon distribution which went negative for $x>0.4$ due to lack of any data constraint.  This implies a negative charm structure function, $F_2^{\rm charm}$, and a terrible description ($\chi^2/N_{\rm pts.}\sim 10$ including correlated systematic errors) of Tevatron jet data using the obtained PDFs.  A DIS-only fit fixing the high-$x$ gluon parameters to prevent such bad behaviour gave $\alpha_S(M_Z^2) = 0.1172$, i.e.~very similar to the global fit.  However, a NNLO fit which imposed the condition of the positive low-$x$ gluon, which stopped the gluon from going negative at high $x$ values, and which also omitted the Tevatron jet data, gave $\alpha_S(M_Z^2)=0.1139$, rather closer to the ABKM09 value.  The very low value of $\alpha_S(M_Z^2) = 0.1104$ found in the DIS-only fit is due to the dominance of BCDMS data.  We can show this explicitly by removing the BCDMS data from the DIS-only fit, then the best-fit $\alpha_S(M_Z^2)$ moves from $0.1104$ to $0.1193$.  Repeating the \emph{global} fit with BCDMS data removed gives $\alpha_S(M_Z^2) = 0.1181$, i.e.~a change by less than the quoted experimental uncertainty of $\pm0.0014$.  The conclusion is that the Tevatron jet data are vital to pin down the high-$x$ gluon, giving a smaller low-$x$ gluon and therefore a larger $\alpha_S$ in the global fit compared to a DIS-only fit, at the expense of some deterioration in the fit quality of the BCDMS data.  The benefits of including the Tevatron jet data to obtain sensible results in a simultaneous fit of PDFs and $\alpha_S$ therefore greatly outweighs any disadvantage such as lack of complete NNLO corrections.

The only input DIS value to the current world average $\alpha_S(M_Z^2)$~\cite{Bethke:2009jm,Nakamura:2010zzi} is the BBG06 value~\cite{Blumlein:2006be}, which is from a non-singlet analysis and therefore \emph{in principle} free of assumptions made about the gluon distribution.  A value of
\[
  \hspace*{-5mm}\alpha_S(M_Z^2)=\left\{0.1148^{+0.0019}_{-0.0019}, 0.1134^{+0.0019}_{-0.0021}, 0.1141^{+0.0020}_{-0.0022}\right\}
\]
was obtained at \{NLO, NNLO, N$^{3}$LO\}, by fitting proton and deuteron structure functions, $F_2^p$ and $F_2^d$, for $x\ge 0.3$ (assuming only valence quarks, neglecting the singlet contribution), and the less precise $F_2^{\rm NS}=2(F_2^p-F_2^d)$ for $x<0.3$.  However, using the MSTW08 NNLO central fit, contributions other than valence quarks are found to make up about $10\%$ ($2\%$) of $F_2^p$ at $x=0.3$ ($x=0.5$).  As an exercise we performed the MSTW08 NNLO DIS-only fit just to $F_2^p$ and $F_2^d$ for $x>0.3$ (comprising 282 data points, 160 of these from BCDMS), which gave $\alpha_S(M_Z^2)=0.1103$ ($0.1130$) without (with) the singlet contribution included.  This is even lower than the BBG06 value presumably due to lack of the $y>0.3$ cut on BCDMS data applied in the BBG06 analysis.  The low value of $\alpha_S(M_Z^2)$ found by BBG06~\cite{Blumlein:2006be} is therefore due to both dominance of BCDMS data and by what we conclude is the unjustified neglect of the singlet contribution to $F_2^p$ and $F_2^d$ for $x\ge 0.3$.  Given that it was argued above that the Tevatron jet data are needed to pin down the high-$x$ gluon, we conclude that an extraction of $\alpha_S(M_Z^2)$ only from inclusive DIS data is not meaningful, and the closest possible to a reliable extraction is the MSTW08 NNLO combined analysis of DIS, Drell--Yan and jet data~\cite{Martin:2009iq,Martin:2009bu},
\[
  \alpha_S(M_Z^2) = 0.1171\pm0.0014~{\rm(exp.)\pm0.002~{\rm(th.)}},
\]
or, alternatively, the more recent NNLO determination by the NNPDF Collaboration~\cite{Ball:2011us},
\[
\alpha_S(M_Z^2) = 0.1173\pm0.0007~{\rm(exp.)\pm0.0009~{\rm(th.)}}.
\]
These values are the only NNLO determinations, from a simultaneous fit with PDFs, which are in agreement with the current world average $\alpha_S(M_Z^2)=0.1184\pm0.0007$~\cite{Bethke:2009jm,Nakamura:2010zzi}; see Fig.~\ref{fig:alphaSMZ}.

With all these problems in $\alpha_S$ determinations from non-global PDF fits, it is therefore disconcerting that the 2011 update of the world average $\alpha_S$ by S.~Bethke~\cite{Bethke:2011}, intended for the PDG 2012 review, has chosen to treat the MSTW08, ABKM09, JR09 and BBG06 determinations on an equal footing in forming a ``DIS'' value of $\alpha_S(M_Z^2)=0.1148\pm0.0024$, which lies significantly below other classes of measurements and the overall preliminary world average of $\alpha_S(M_Z^2)=0.1185\pm0.0008$~\cite{Bethke:2011}.

\section{Summary}
\label{sec:summary}

The ``MSTW 2008'' determination of parton distribution functions (PDFs)~\cite{Martin:2009iq,Martin:2009bu,Martin:2010db} is still fairly current with no immediate update planned.  Another update may be appropriate after the final HERA II combined data (including $F_2^{\rm charm}$) are published, together with making the first use of input data from the LHC, such as differential $Z/\gamma^*$, $W$, jet and $W$+charm distributions.  However, provided that the new data are reasonably consistent with the old data, we would not expect to see substantial changes to the PDFs.  On the other hand, perhaps a more pressing concern is the compatibility of existing PDFs from different groups with each other.  We have updated a recently-published study of benchmark cross sections~\cite{Watt:2011kp,Thorne:2011kq} to include results obtained using the NNPDF2.1 NNLO~\cite{Ball:2011uy} and HERAPDF1.5 NNLO~\cite{HERA:2011} PDF sets, and to compare to LHC data on $W$, $Z$ and $t\bar{t}$ production.  Supplementary plots can be obtained from a public webpage~\cite{mstwpdf}.  There is now reasonably good agreement between the NLO \emph{global} fits from MSTW08, CT10 and NNPDF2.1, all using variants of a GM-VFNS to treat DIS structure functions, together with the NNLO \emph{global} fits from MSTW08 and NNPDF2.1.  More variation is seen with other PDF sets using more limited data sets and/or restrictive input PDF parameterisations.  The latest HERAPDF1.5 NNLO set is surprisingly close to MSTW08, at least for the central value, unlike the analogous HERAPDF1.5 NLO set, but it has a large uncertainty in the high-$x$ gluon distribution due to variation of the minimum $Q^2$ cut, not seen at NLO perhaps due to the more restricted parameterisation.  The Tevatron jet data are important to pin down the high-$x$ gluon, with an indirect effect on the value of $\alpha_S(M_Z^2)$ extracted, reducing sensitivity to the fine details of the treatment of DIS data.  The LHC measurements of top-pair production tend to favour PDF sets where the high-$x$ gluon is determined using Tevatron jet data.  Conversely, both the LHC $t\bar{t}$ cross sections and the Tevatron jet data strongly disfavour PDF sets with soft high-$x$ gluon distributions and low $\alpha_S$ values (specifically, ABKM09~\cite{Alekhin:2009ni}), which give anomalously low Higgs cross sections at both the Tevatron and LHC.  We therefore caution against the use of PDF sets obtained from ``non-global'' fits where the high-$x$ gluon distribution is generally constrained by assumptions made on the form of parameterisation in the absence of a direct data constraint.

\nocite{*}
\bibliographystyle{elsarticle-num}
\bibliography{watt}


\end{document}